%% file: ms.tex
\documentclass[final]{elsarticle}
\usepackage[utf8]{inputenc}

\usepackage{grffile}
\usepackage{lineno}
\usepackage{times}
\usepackage{calc}
\usepackage{graphics}
\usepackage{epsfig}
\usepackage{amssymb}
\usepackage{amsmath}
\usepackage{flafter}
\usepackage{float}
\usepackage{latexsym}
\usepackage{psfrag}
\usepackage{dsfont}
\usepackage{rotating}
\usepackage{longtable}
\usepackage{cleveref}
\usepackage{color}
\usepackage{longtable,lscape}
\usepackage{multirow}
\usepackage{tabularx}
\usepackage{booktabs}
\usepackage{subcaption}
\usepackage{array}
\usepackage{soul}
\newcolumntype{M}[1]{>{\centering\arraybackslash}m{#1}}
\usepackage{pgfplots}
\usepackage{tikzscale}
\usepackage{pgfkeys}
\usepackage{pgfplotstable}
\usetikzlibrary{matrix}
\usepgfplotslibrary{groupplots}
\pgfplotsset{compat=newest}
\pgfplotsset{}
\pgfplotsset{
	tick label style={font=\footnotesize},
	label style={font=\footnotesize},
	legend style={font=\footnotesize}
}
\pgfplotsset{major grid style={loosely dotted, thin, gray}}
\pgfkeys{
	/includetikzgraphic/.is family, /includetikzgraphic,
	default/.style = {
		width = 0.85\textwidth,
		ratio = 0.85},
	width/.estore in = \itgWidth,
	ratio/.estore in = \itgRatio
}
\newlength\figurewidth
\newlength\figureheight

\modulolinenumbers[5]

\journal{Journal of \LaTeX\ Templates}

\newtheorem{remark}{Remark}

\newcommand{\figref}[1]{Figure~\ref{#1}}
\newcommand{\tabref}[1]{Table~\ref{#1}}
\newcommand{\secref}[1]{Section~\ref{#1}}
\newcommand{\rmkref}[1]{Remark~\ref{#1}}
\LetLtxMacro{\originaleqref}{\eqref} % Save command, maybe needed at some point
\renewcommand{\eqref}{equation~\originaleqref}

\begin{document}
	
\begin{frontmatter}

\title{A machine learning framework for LES closure terms}

\author{Marius Kurz}
\author{Andrea D. Beck}

\address{Institute of Aerodynamics and Gas Dynamics, University of Stuttgart, Stuttgart, Germany}

\begin{abstract}
  In the present work, we explore the capability of artificial neural networks~(ANN) to predict the closure terms for large eddy simulations~(LES) solely from coarse-scale data.
  To this end, we derive a consistent framework for LES closure models, with special emphasis laid upon the incorporation of implicit discretization-based filters and numerical approximation errors.
  We investigate implicit filter types, which are inspired by the solution representation of discontinuous Galerkin and finite volume schemes and mimic the behaviour of the discretization operator, and a global Fourier cutoff filter as a representative of a typical explicit LES filter.
  Within the perfect LES framework, we compute the exact closure terms for the different LES filter functions from direct numerical simulation results of decaying homogeneous isotropic turbulence.
  Multiple ANN with a multilayer perceptron~(MLP) or a gated recurrent unit~(GRU) architecture are trained to predict the computed closure terms solely from coarse-scale input data.
  For the given application, the GRU architecture clearly outperforms the MLP networks in terms of accuracy, whilst reaching up to~$99.9\%$ cross-correlation between the networks' predictions and the exact closure terms for all considered filter functions.
  The GRU networks are also shown to generalize well across different LES filters and resolutions.
  The present study can thus be seen as a starting point for the investigation of data-based modeling approaches for LES, which not only include the physical closure terms, but account for the discretization effects in implicitly filtered LES as well.  
\end{abstract}
\end{frontmatter}

% Main Part
\input{input_tex/1_introduction}
\input{input_tex/2_les}
\input{input_tex/3_ann}
\input{input_tex/4_results}
\input{input_tex/5_conclusion}

\section*{Acknowledgments}
This research was funded by Deutsche Forschungsgemeinschaft (DFG, German ResearchFoundation) under Germany's Excellence Strategy - EXC 2075 - 390740016.
The authors gratefully acknowledge the support and the computing time on ``Hazel Hen'' provided by the HLRS through the project ``hpcdg''.

\bibliography{ms.bib}{}
\bibliographystyle{siamplain}

\end{document}

%% file: input_tex/1_introduction.tex
\section{Introduction}
Over the last decade, machine learning methods, and in particular artificial neural networks~(ANN), have achieved tremendous success: from pushing the state-of-the-art in the fields of image and speech recognition~\cite{Hinton2012,LeCun2015} to surpassing human-level performance in the game of Go~\cite{Silver2016,Silver2018}.
This recent success of ANN was predominantly fuelled by the emergence of large datasets, the exploitation of highly-parallel graphical processing units (GPU)~\cite{Hinton2012}, and the development of easy-to-use high-performance machine learning libraries like Pytorch~\cite{Pytorch2019} and Tensorflow~\cite{Tensorflow2015}.
In general, artificial neural networks can approximate any continuous functional relationship between input and output quantities solely based on data and without prior assumptions on the nature of said function.
These approximation properties are shown by a variety of universal approximation theorems in literature, e.g., \cite{Cybenko1989,Hornik1991,Lu2017}.\\
Significant efforts have been directed towards utilizing the approximation capabilities of neural networks also for problems in other scientific fields, including the field of turbulence research.
Turbulence is a multi-scale phenomenon, for which the range of active scales in the flow grows with increasing Reynolds numbers.
Thus, direct numerical simulation~(DNS) at high Reynolds numbers is still computationally prohibitive for most applications.
Common approaches to mitigate the computational effort are the Reynolds-averaged Navier-Stokes equations~(RANS) or the method of large eddy simulation~(LES), which both rely on filter operations to resolve only the most influential part of the solution, while modeling the effects of the non-resolved part.
The filtering of the non-linear convective terms of the Navier-Stokes equations evokes the so-called closure problem, which introduces an additional model term in the coarse-scale equations.
For the RANS approach, only the temporally averaged solution is resolved and the influence of the turbulent fluctuations is modelled, while the LES approach resolves the large, energy-containing flow scales in space and time, but uses models to describe the non-resolved fine scale contributions.
These models are typically derived based on physical or mathematical considerations.
Despite decades of research, no model proved itself superior for all applications and many models comprise empirical parameters, which have to be tuned for the respective flow regime and/or the discretization choices~\cite{Sagaut2009}.\\
Recently, increasing efforts have been made to complement the established physics-based metholodogy of turbulence modeling by a data-driven approach based on ANN.
In one of the first works employing neural networks for LES modeling, Sarghini~et~al.~\cite{Sarghini2003} used a neural network to approximate a mixed LES model to save computation time.
A similar approach for RANS was investigated in~\cite{Tracey2015}. Ling~et~al.~\cite{Ling2016} derived a novel ANN architecture to embed physical invariances in the predicted RANS closure terms.
In context of LES, Maulik and San~\cite{Maulik2017} used ANN for approximate deconvolution of filtered turbulence.
The direct inference of the LES subgrid stresses is investigated in~\cite{Maulik2019} for two-dimensional Kraichnan turbulence and in \cite{Gamahara2017} for turbulent channel flow.
Different network architectures were used by Beck~et~al. in~\cite{Beck2019}, who used convolutional neural networks to predict the closure terms for homogeneous isotropic turbulence (HIT), and
Srinivasan~et~al.~\cite{Schlatter2019}, who used long short-term memory networks to predict the turbulent dynamics by means of a shear flow model.
In~\cite{Xie2019}, ANN are used to predict the closure for compressible isotropic turbulence.
More recently, Novati~et~al.~\cite{Novati2020} used a multi-agent reinforcement learning approach to infer local LES model parameters for the HIT test case.\\
In this work, we derive a general machine learning framework for LES closure models, while laying emphasis on discretization effects and the influence of different LES filter forms.
We generate training data by applying several distinct LES filter funtions to DNS data of decaying homogeneous isotropic turbulence (DHIT).
For this, we project the DNS solution onto a discontinuous Galerkin (DG) or finite volume (FV) flavored representation on a coarsened mesh. This resembles the implicitly filtered LES approach, where the discretization acts as an implicit LES filter.
In addition, a global Fourier cutoff filter is used, while ensuring that the filtered solution is perfectly resolved by the underlying numerical scheme to mitigate any influences of the LES discretization.
We show that the multilayer perceptron (MLP) and recurrent neural networks with gated recurrent units (GRU) are generally able to learn the unknown closure terms from data for all filter forms, with the GRU achieving excellent accuracy with up to $99.9\%$ cross-correlation between the predicted and the exact closure terms.
Further, we show that the trained GRU networks generalize well across different LES resolutions and different filter functions.
This is a surprising and encouraging result, since an accurate prediction of the actual closure terms would greatly benefit the development and help guide the selection of proper LES closure models.\\
This paper is organized as follows:
In \secref{sec:les}, the underlying equations are discussed and special emphasis is laid upon the derivation of the perfect LES closure terms.
The relevant network architectures are discussed in \secref{sec:ann} and the training of the networks is described in \secref{sec:training}.
The results are reported in \secref{sec:results}. \secref{sec:conclusion} concludes the paper.

%% file: input_tex/2_les.tex
\section{Turbulence modelling}
\label{sec:les}
Since neural networks approximate functional relationships solely from data, special care has to be taken to ensure its validity and consistency.
Based on the compressible Navier-Stokes equations in \secref{sec:nse}, the LES closure terms are derived in a consistent manner in \secref{sec:perfect_les} as the overall target quantity of the machine learning task.
For the DHIT test case described in \secref{sec:test_case}, these closure terms are then computed and examined in \secref{sec:closure_terms}.

\newcommand{\ppvector}[1]{#1}
\newcommand{\ppmatrix}[1]{\ensuremath{\underline{#1}}}
\newcommand{\phyflux}{\ensuremath{F}}
\newcommand{\flux}{\ensuremath{\mathcal{F}}}
\newcommand{\inviscid}[1]{\ensuremath{#1^{c}}}
\newcommand{\viscous}[1]{\ensuremath{#1^{v}}}

\subsection{Governing equations}
\label{sec:nse}
The Navier-Stokes equations describe the evolution of compressible, viscous fluids and can be written in conservative form as
\begin{equation}
	\ppvector{U}_t + \ppvector{\nabla}_x \cdot \inviscid{\ppvector{\phyflux}}(\ppvector{U}) - \ppvector{\nabla}_x \cdot \viscous{\ppvector{\phyflux}}\left(\ppvector{U},\ppvector{\nabla}_x \ppvector{U} \right) = 0\;.
  \label{eq:navier_stokes}
\end{equation}
The vector of conserved variables is $\ppvector{U} = \left[\rho,\rho v_1,\rho v_2,\rho v_3, \rho e\right]^T$ with mass, momentum und energy, respectively. $\ppvector{U}_t$ denotes differentation with respect to time and the differential operator with respect to the spatial coordinates is denoted as~$\ppvector{\nabla}_x$. The convective fluxes \inviscid{\ppvector{\phyflux}} and the viscous fluxes \viscous{\ppvector{\phyflux}} with columns $i=1,2,3$ take the form
\begin{equation}
  \inviscid{\ppvector{\phyflux}}_i =
  \left[
  \begin{array}{c}
    \rho v_i \\
    \rho v_1 v_i + \delta_{1i} p\\
    \rho v_2 v_i + \delta_{2i} p\\
    \rho v_3 v_i + \delta_{3i} p\\
    \rho e v_i + p v_i
  \end{array}
  \right]
  ,\;
  \viscous{\ppvector{\phyflux}}_i =
  \left[
  \begin{array}{c}
    0 \\
    \tau_{1i}\\
    \tau_{2i}\\
    \tau_{3i}\\
    \tau_{ij}v_j-q_i
  \end{array}
  \right] \;,
  \label{eq:fluxes_written_out}
\end{equation}
with $\delta$ denoting the Kronecker delta and $p$ as static pressure. The stress tensor~$\tau_{ij}$ and the heat flux~$q_i$ can be written as
\begin{align}
  \tau_{ij} &= \mu\left(\frac{\partial v_i}{\partial x_j}+\frac{\partial v_j}{\partial x_i}-\frac{2}{3}\delta_{ij}\frac{\partial v_k}{\partial x_k}\right) \;,
  \label{eq:stress_tensor}\\
  q_i &= - k\:\frac{\partial T}{\partial x_i}    \;,
  \label{eq:heat_flux}
\end{align}
where $k$ denotes the heat conductivity, $T$ is the static temperature, and $\mu$ is the dynamic viscosity. Assuming an ideal gas, the equation system is closed by the relation
\begin{equation}
 p = \rho\left(\gamma-1\right) \left[e-\frac{1}{2}\left(v_1^2+v_2^2+v_3^2\right)\right]\;,
\end{equation}
with $\gamma$ as ratio of specific heats.
In this work, the compressible equations are only solved for small Mach numbers, i.e., in the limit $\mathrm{Ma}\rightarrow0$, for which compressibility effects become negligible and eventually the incompressible Navier-Stokes equations are solved.

\subsection{Perfect large eddy simulation}
\label{sec:perfect_les}
In numerical simulations, the continuous form of \eqref{eq:navier_stokes} is replaced by a discretized formulation, which can for instance be written as
\begin{equation}
  U_t + R(F(U)) = 0 \;,
  \label{eq:discrete_nse}
\end{equation}
with $R$ denoting a discrete divergence operator applied to the fluxes $F=F^c-F^v$.
It is clear that the discretization operator is consistent in the limit $h\rightarrow0$, i.e., if \eqref{eq:discrete_nse} is solved on a grid with negligible discretization errors, the continuous solution $U$ is recovered.
For turbulent problems, \eqref{eq:discrete_nse} often cannot be solved on such a fine grid, due to the multi-scale character of turbulence and the consequently prohibitive computational cost.
The methodology of large eddy simulation (LES) avoids to resolve the expensive fine-scale components of the solution by only resolving its coarse scales.
This corresponds to applying a low-pass filter~$\overline{()}$ with linear filter kernel to \eqref{eq:discrete_nse}.
Under the common assumption that filter and time derivative commute, this yields
\begin{equation}
  \overline{U}_t + \overline{R(F(U))} = 0 %\;,
  \label{eq:filtered_nse}
\end{equation}
as the exact evolution equation for the coarse-scale solution~$\overline{U}$.
Note that, although \eqref{eq:filtered_nse} is a coarse-scale formulation, the closure problem has not been solved yet: Computation of the second term in \eqref{eq:filtered_nse} would require the knowledge of $U$ (on a DNS level), since $F$ depends non-linearly on it. 
For practical LES,~$\overline{U}$ is advanced in time using an appropriate numerical scheme with its spatial discretization operator~$\tilde{R}(\overline{U})$ applied to the coarse-scale solution.
Note that $\tilde{R}\rightarrow R$ in the above limit of $h\rightarrow0$, but that $\tilde{R}$ introduces approximation errors otherwise.
Further, $\tilde{R}$ can be non-linear. In fact, discretization operators typically contain some form of non-linearity to ensure stability in underresolved regions. We use this notation to highlight that while the filter operator $\left(\bar{.}\right)$ is known analytically, the discretization induced filtering $\left(\tilde{.}\right)$ is not.
Equation~\originaleqref{eq:filtered_nse} can then be written as
\begin{equation}
  \overline{U}_t + \tilde{R}(\overline{U}) = \underbrace{\tilde{R}(\overline{U}) - \overline{R(F(U))}}_{\mathrm{perfect\:LES\:closure}}\;.
  \label{eq:perfect_les}
\end{equation}
The described approach yields the perfect LES closure term as the right-hand side of \eqref{eq:perfect_les}\footnote{Note that \eqref{eq:perfect_les} is written with the practical application of an implicitly filtered LES in mind. The explicitly filtered LES formulation can be derived by applying an additional $\left(\bar{.}\right)$ filter to \eqref{eq:perfect_les}. In this case, grid refinement under the filter will ensure $\tilde{R}\rightarrow\bar{R} $.} .
If the closure term~$(\tilde{R}(\overline{U})-\overline{R(F(U))})$ is known during simulation (e.g., computed from high-fidelity DNS data), the \textit{exact} coarse-scale solution $\overline{U}$ can be recovered, as is verified in \cite{Beck2019}.
Since this DNS data is not available in practical situtions, the right-hand side of \eqref{eq:perfect_les} is typically replaced by a model, which is based on physical or mathematical reasoning, to solve this closure problem.
A more detailed discussion on the approach of \textit{perfect LES} can be found in~\cite{Beck2019}, we merely wish to emphasize here that this is a formal framework for the analysis of LES.\\
By establishing \eqref{eq:perfect_les}, two distinct choices have to be made. The first is the choice of the LES operator.
Since the LES discretization is a constituent part of the perfect closure terms, approximate closure models with a given set of model parameters generally cannot be expected to be optimal for a wide variety of discretizations. An appropriate closure model and its parameters rather have to be chosen discretization-specific by design.
The second and more obvious choice is the selection of the LES filter, which directly defines the coarse-scale space and the unknown component of the closure term $\overline{R(F(U))}$. More importantly, the chosen filter function also predefines the coarse-scale solution $\overline{U}$ itself, and thus the overall target quantity of LES. Altering the LES filter therefore not only changes the closure terms, but also the exact coarse-scale solution.
\begin{remark}
  Important to stress is that, if the operator $\tilde{R}$ is linear, the distinction between filtering the discrete or the continuous equations becomes redundant, since in this special case the \textit{linear} filter and the \textit{linear} operator do, in fact, commute (disregarding possible issues with violation of homogeneity on stretched grids). This is also typically the case, if the coarse-scale solution is perfectly resolved by the numerical scheme and the closure term does not introduce subfilter scales, since then the numerical approximation error becomes negligible. Under these assumptions, also the discretization-dependence of approximate closure models vanishes.
Note that this special case is labelled \textit{explicitly filtered} LES. Here, the scale separation filter and the discretization are completely independent from each other, and spatial convergence of the filtered solution is simply achieved by grid refinement.
  %\mk{Formally, this can be written as an additional low pass filtering of \eqref{eq:perfect_les}}.
  The opposite approach is the \textit{implicitly filtered} LES, in which discretization and filter are intrinsically linked and the discretization operator defines the coarse-scale solution.
  This has tremendous advantages in terms of computational efficiency, but makes analysis very tough.
Our chosen approach of the perfect LES is an attempt to improve this situation.
  We recognize that some (unknown) filtering is introduced by the numerical scheme, which defines the expected solution and the closure terms.
  Note that we cannot specify a filter kernel, but we can make educated guesses for what the kernel might look like, as is discussed in \secref{sec:closure_terms}.
  We also incorporate the fact that we do know that spatial discretization is applied, i.e., we consider the effects of a non-errorfree operator $\tilde{R}$ in \secref{sec:closure_terms}.
  \label{rmk:nonlinear_operator}
\end{remark}

\subsection{Homogeneous isotropic turbulence}
\label{sec:test_case}
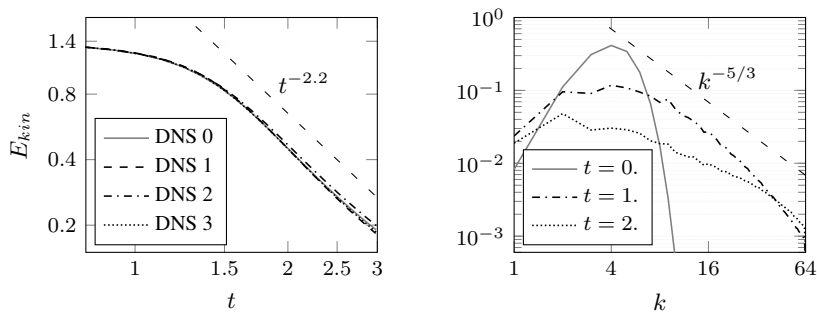
\begin{figure}[htb!]
  \centering
  \input{./tikz/fig_ekin.tikz}
  \caption{The temporal evolution of the integral kinetic energy over time for several DNS of the DHIT test case (\textit{left}) and the distribution of kinetic energy over the wavenumbers $k$ at distinct time instants for a single DNS (\textit{right}).}
  \label{fig:ekin}
\end{figure}

To investigate the nature of the LES closure, it can be exploited that the perfect closure terms can be computed exactly from high-fidelity DNS data, as indicated in \secref{sec:perfect_les}. To this end, decaying homogeneous isotropic turbulence~(DHIT) is investigated, which is a common test case for LES closure models and is studied extensively in literature, e.g., \cite{Sagaut2018}. For this test case, a periodic box is initialized with a pseudo-turbulent flow field. The transfer of kinetic energy from lower to higher wavenumbers and the viscous dissipation then lead to a decay of kinetic energy over time, as is shown in \figref{fig:ekin}.
In this work, random pseudo-turbulent flow field realizations for a prescribed energy spectrum proposed by Chasnov~\cite{Chasnov1995} are used as initial conditions for the DHIT simulations. The flow fields are obtained with the procedure proposed by Rogallo in \cite{Rogallo1981} for incompressible flows. To obtain the compressible state from the incompressible velocity field, the mean pressure of the initial flow field is chosen in order to obtain a Mach number of 0.1 with respect to the maximum velocity in the initial flow field.
The pressure fluctuations are then set thermodynamically consistent to the prescribed flow field with unit density. The Reynolds number with respect to the Taylor microscale is approximately 180.
All time specifications in the following are normalized by unit length and unit velocity.
This test case of ``turbulence in a box'' can be seen as the building block of turbulence in the absence of boundary conditions and is thus highly suitable to study turbulent dynamics and model development.
\\
% Computation
The DNS of the test case were conducted with the discontinuous Galerkin solver FLEXI~\cite{Krais2020} on the supercomputer Hazel Hen at the High-Performance Computing Center Stuttgart (HLRS). An eighth-order DG scheme was used on a Cartesian mesh with $64^3$~elements, yielding around 134~million~degrees of freedom per solution variable. The domain is of size $[0,2\pi]^3$ with periodic boundary conditions. The computational cost for a single DNS up to $t=3$ was around $15,000$~CPU-hours. In addition to the solution~$U$, also the term~$R(F(U))$ was stored during the simulation.
\subsection{The exact closure terms}
\label{sec:closure_terms}
The DNS solution is filtered in a post-processing step to obtain the coarse-scale quantities, i.e., the coarse-scale solution~$\overline{U}$ and the filtered DNS divergence~$\overline{R(F(U))}$.
Three distinct filter shapes are investigated in this work.
As discussed above, the actual filter that corresponds to a discretization is unkown and typically non-linear and inhomogeneous, thus, we choose filter forms associated with typical discretization operators.
The first filter is a local $L_2$-projection onto piecewise polynomials, which arises naturally from the discontinuous Galerkin~(DG) framework. To this end, the DNS solution is projected element-wise on a coarse LES grid with $8^3$ equi-spaced elements. In each of these elements, the solution is approximated by a polynomial basis of polynomial degree $N=5$ on Gauss-Lobatto interpolation points, which is a common solution representation for DG methods.
This results in 48 degrees of freedom in every spatial direction.
This filter can be seen as an approximation to the true DG filter function in the limit for $N\rightarrow\infty$, without regarding the interface coupling.
% Filter energy spectra
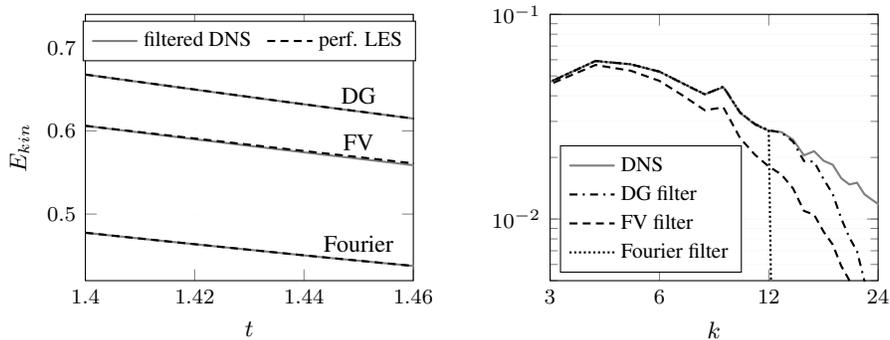
\begin{figure}[htb!]
  \centering
  \input{./tikz/fig_spectra_filter.tikz}
  \caption{Shown on the left is the temporal evolution of kinetic energy for the filtered DNS and the perfect LES for the FV, DG, and Fourier filter, respectively. Shown on the right are the energy spectra at $t=1.4$ for the DNS solution and the different LES filters.}
  \label{fig:spectra_filter}
\end{figure}\\
Secondly, the DNS solution is filtered by an $L_2$-projection onto a piecewise constant representation on a Cartesian mesh with $48^3$ elements, again resulting in 48 degrees of freedom in every spatial direction. This filter can be interpreted as a finite volume~(FV) flavored representation of the solution. In contrast to the DG filter, this representation is only first-order accurate.
Due to the reduced number of points per wavelength resolution capability \cite{Gassner2013}, the spectral distribution of kinetic energy in the coarse-scale solution exhibits greater deviations from the full DNS solution than the DG representation does, as shown in \figref{fig:spectra_filter}.\\
In contrast to the first two local LES filters, the third examined LES filter is a global Fourier cutoff filter.
To this end, a Fourier transform is applied to the DNS solution, followed by a cutoff filter at wavelength $k_{max}$. The DG representation discussed above is then used to represent the filtered solution again in physical space. The cutoff wavelength is chosen as $k_{max}=12$. This ensures that all wavelengths present in the solution are represented accurately by the underlying DG representation, as can be seen in \figref{fig:spectra_filter}.
This choice of a globally defined filter and a discretization that resolves the occuring scales with negligible error corresponds to the case of the explicitly filtered LES described above.
\begin{figure}[htb!]
  \centering
  \input{./tikz/fig_filter_field_solution.tikz}
  \caption{Two-dimensional slices of the full and the filtered x-velocity field~(\textit{top}) and flux divergence~(\textit{bottom}) at~$t=1.4$. Shown is the field solution (from left to right) for the DNS, the local DG filter, the local FV filter, and the global Fourier filter.}
  \label{fig:filter_field_solution}
\end{figure}
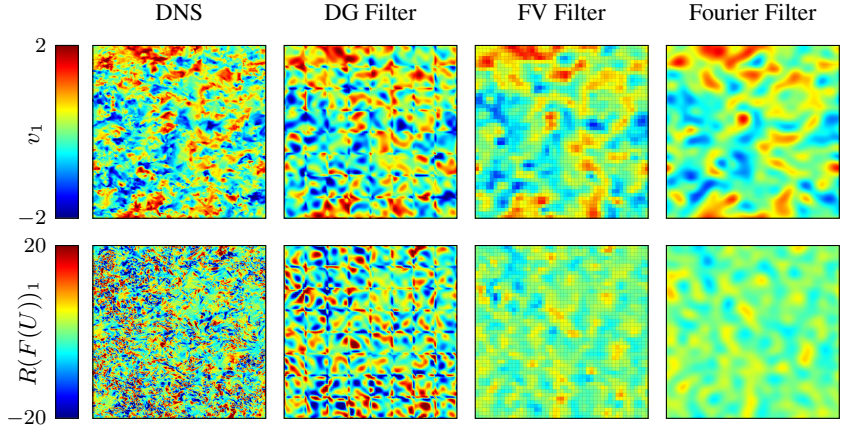\\
The three chosen filter types thus cover the range of typically used LES methods: from the global filtering associated with pseudospectral methods to local filter operations onto a polynomial subspace with the FV discretization as the limit in terms of locality.
The full and the filtered field solution of the x-velocity $v_1$ and the DNS operator~$R(F(U))_1$ are shown in \figref{fig:filter_field_solution} for all three filter forms.
It is important to stress that only the Fourier filter is strictly homogeneous, for which Moser~et~al.~\cite{Moser2021} argue that discretization effects do not have to be modeled, since differentiation and filtering commute.\\
It was verified that the derived closure terms are indeed perfect in the sense that they recover the filtered DNS solution in every timestep. To this end, the DNS results were filtered in the interval $t\in [1.40,1.46]$ at each LES timestep $\Delta t=10^{-4}$.
To compute the perfect closure terms according to \eqref{eq:perfect_les}, the described sixth-order DG scheme is used as LES discretization for the DG- and Fourier-filtered data, while a first-order FV scheme ist used for the FV data.
These computed exact closure terms are then used as source terms during an LES with the respective spatial discretization scheme. A third-order Adams-Bashforth method with a fixed timestep of $\Delta t=10^{-4}$ is used for time integration, to ensure that the perfect closure terms are known in each timestep of the simulation.
The results of these simulations in \figref{fig:filter_field_solution} demonstrate that the obtained coarse-scale solution coincides with the respectively filtered DNS solution for all considered filter forms.
This confirms our definition of a \textit{perfect} LES. In its own right,~\figref{fig:filter_field_solution} already reveals some interesting insights. Firstly, as expected, the choice of the filter $\left(\bar{.}\right)$ defines the resolved field $\bar{U}$ as a coarse scale representation of the full solution. The resolved fields show different properties in terms of locality and smoothness as caused by the filter. Even more striking however is the difference in the closure terms induced by the filter, underlining the statement that the optimal closure is a direct function of the LES filter form (and thus the discretization operator and its properties). 
\begin{figure}[htb!]
  \centering
  \input{./tikz/fig_filter_histogramm.tikz}
  \caption{Histograms of the magnitude of the full closure terms at $t=1.4$. Shown are the distributions of the DG filter, the FV filter, and the Fourier filter for the closure terms of the mass equation~(\textit{left}), the x-momentum equation~(\textit{center}), and the energy equation~(\textit{right}). The histograms are normalized to integrate to unity.}
  \label{fig:filter_histogramm}
\end{figure}
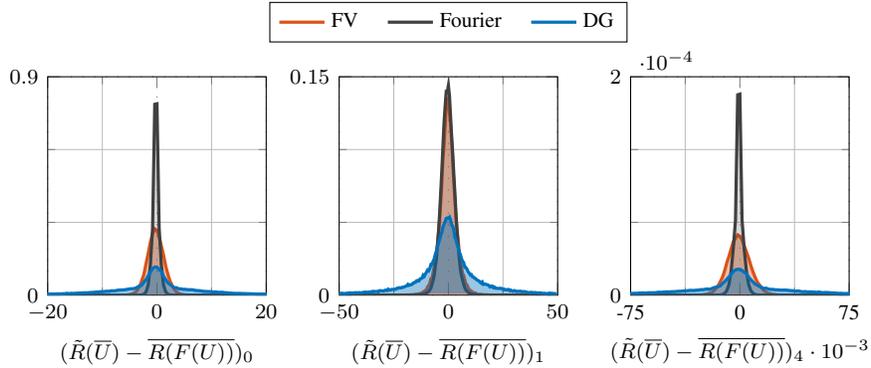\\
Supporting this, further examination of the closure terms reveals that their statistics strongly depend on the chosen LES filter, as shown in \figref{fig:filter_histogramm}.
While the Fourier and FV filter show similar distributions of the closure terms in the momentum equation, the distribution of the DG-filtered data exhibits significantly higher variance.
This is caused by the element-wise $L_2$-projection, which leads to large discontinuities in the solution at the element interfaces, c.f.~\figref{fig:filter_field_solution}.
These discontinuities lead to large flux contributions by the interface coupling and therefore to an overall increase in magnitude for the closure terms.
This again stresses the influence of the discretization on the LES closure terms for the implicitly filtered LES approach.
\begin{remark}
  A common assumption for low Mach number flows is that the closure terms of the mass and energy equation become negligible in comparison to the momentum closure, see e.g.~\cite{Sagaut2009}.
While this assumption is typically derived in context of the continuous formulation, the assumption might be invalid if discretization effects must be taken into account.
As shown in \figref{fig:filter_histogramm}, while the closure terms for the Fourier filtered approach for the mass and energy equations are generally very close to zero as denoted by the sharp peak, this is neither for the DG nor for the FV filter the case, which feature rather wide tails. Interestingly, the momentum closure terms for the FV and Fourier filter are similar in shape, while the DG data again shows a much wider distribution. We attribute this behaviour to the hybrid nature of the DG operator and the high order approximation used here. Our analysis thus indicates that the LES filter (and the discretization scheme) does influence the validity of the discussed assumption.
  This notion is also supported by Moser~et~al.~\cite{Moser2021}. In order to test these propositions, the perfect LES computations from \figref{fig:filter_field_solution} were repeated, but instead of closing all five equations, only the momentum equations are closed in the perfect LES, while leaving the mass and energy equation unclosed.
For the Fourier filter, the results showed only negligible deviations from the previous computation, and still matched the filtered DNS data, as expected for the explicitly filtered approach. In contrast, the solution for the DG filter diverged quickly and irrefutably from the filtered DNS data. 
This demonstrates that the assumption, that the closure terms in mass and energy equation are negligible for low Mach number flows does hold for explicitly filtered LES, but not necessarily for implicitly filtered LES, where discretization effects become important.
  \label{rmk:closure_mass_energy}
\end{remark}

%% file: tikz/fig_ekin.tikz
\pgfplotscreateplotcyclelist{mycolorlist_t}{%
semithick, gray,  solid,  mark=none \\%
semithick, black, dashed, mark=none \\%
semithick, black, dashdotted, mark=none \\%
semithick, black, densely dotted, mark=none \\%
}
\pgfplotscreateplotcyclelist{mycolorlist_k}{%
semithick, gray,  solid,  mark=none \\%
semithick, black, dashdotted, mark=none \\%
semithick, black, densely dotted, mark=none \\%
%black, denselydotted, mark=none \\%
}
\begin{tikzpicture}[font=\small,baseline]
  \begin{groupplot}[
        group style={
          group size=2 by 1,
          horizontal sep={0.15\textwidth},
          %vertical sep={0.03\textwidth}
        },
        %width=0.7\linewidth%,height=5cm,
    ]

  \nextgroupplot[
      width={0.45\textwidth},
      grid=both,
      grid style={line width=.1pt, draw=gray!5},
      major grid style={line width=.2pt,draw=gray!25},
      xmin=0.8,
      xmax=3,
      xmode=log,
      xtick={1.0,1.5,2.0,2.5,3.0},
      ymin=0.15,
      ymax=1.8,
      ytick={0.2,0.4,0.8,1.4},
      ymode=log,
      xlabel near ticks,
      ylabel near ticks,
      log ticks with fixed point,
      cycle list name=mycolorlist_t,
      %cycle list name=black white,
      %mark repeat={1000},
      legend pos=south west,
      legend cell align={left},
      legend style={font=\footnotesize},
      xlabel={$t$},
      ylabel={$E_{kin}$}
    ]
  \addplot table [x=Time, y=Ekin comp, col sep=comma, each nth point=200, mark=*, mark repeat=9,mark phase=5] {./data/ekin/Re2000_N7_64Cells_run100_TGVAnalysis.csv};
  \addlegendentry{DNS 0}

  \addplot table [x=Time, y=Ekin comp, col sep=comma, each nth point=200, mark=*, mark repeat=11] {./data/ekin/Re2000_N7_64Cells_run101_TGVAnalysis.csv};
  \addlegendentry{DNS 1}

  \addplot table [x=Time, y=Ekin comp, col sep=comma, each nth point=200, mark=*, mark repeat=12] {./data/ekin/Re2000_N7_64Cells_run102_TGVAnalysis.csv};
  \addlegendentry{DNS 2}

  \addplot table [x=Time, y=Ekin comp, col sep=comma, each nth point=100] {./data/ekin/Re2000_N7_64Cells_run103_TGVAnalysis.csv};
  \addlegendentry{DNS 3}

  %\addplot table [x=Time, y=Ekin comp, col sep=comma, each nth point=100] {./data/ekin/Re2000_N7_64Cells_run104_TGVAnalysis.csv};
  %\addlegendentry{DNS 4}

  \addplot[black, domain=0:3, loosely dashed] (x,{3*x^(-2.2)});

  \draw node[label={[font=\footnotesize]}] at (axis cs:2.15,0.9) {$t^{-2.2}$};

  \nextgroupplot[
      width={0.45\textwidth},
      grid=both,
      grid style={line width=.1pt, draw=gray!5},
      major grid style={line width=.2pt,draw=gray!25},
      xmin=1,
      xmax=64.,
      xmode=log,
      xticklabel=\pgfmathparse{round(exp(\tick))}\pgfmathprintnumber{\pgfmathresult},
      xtick={1.,4.,16.,64.},
      ymin={6*10^(-4)},
      ymax={1.},
      xlabel near ticks,
      ylabel near ticks,
      ymode=log,
      cycle list name=mycolorlist_k,
      legend pos=south west,
      legend style={font=\footnotesize},
      xlabel={$k$},
      legend image post style={mark indices={}},
      mark indices={4,8,16,32}
    ]
  \addplot table [x index=0, y index=1, skip first n=2, col sep=space] {./data/spectra_dns_run100/Re2000_N7_64Cells_run100_EnergySpectrum_0000000.0000000.dat};
  \addlegendentry{$t=0.$}

  \addplot table [x index=0, y index=1, skip first n=2, col sep=space] {./data/spectra_dns_run100/Re2000_N7_64Cells_run100_EnergySpectrum_0000001.0000000.dat};
  \addlegendentry{$t=1.$}

  \addplot table [x index=0, y index=1, skip first n=2, col sep=space] {./data/spectra_dns_run100/Re2000_N7_64Cells_run100_EnergySpectrum_0000002.0000000.dat};
  \addlegendentry{$t=2.$}

  \addplot[black, domain=1:64, loosely dashed] (x,{7.*x^(-5/3)});

  \draw node[label={[font=\footnotesize]}] at (axis cs:21,0.15) {$k^{-5/3}$};

  \end{groupplot}
\end{tikzpicture}

%% file: tikz/fig_spectra_filter.tikz
\pgfplotscreateplotcyclelist{mycolorlist}{%
gray, thick\\%
black, thick, dash dot\\%
black, thick, densely dashed\\%
black, thick, densely dotted\\%
}
\pgfplotscreateplotcyclelist{mycolorlist_t}{%
gray, thick\\%
black, thick, densely dashed\\%
gray, thick\\%
black, thick, densely dashed\\%
gray, thick\\%
black, thick, densely dashed\\%
}

\def\figwidth{0.49\textwidth}

\begin{tikzpicture}[font=\small,baseline]
  \begin{groupplot}[
        group style={
          group size=2 by 1,
          horizontal sep={0.15\textwidth},
          %vertical sep={0.03\textwidth}
        },
        %width=0.7\linewidth%,height=5cm,
    ]
  \nextgroupplot[
      width=\figwidth,
      grid=both,
      grid style={line width=.1pt, draw=gray!5},
      major grid style={line width=.2pt,draw=gray!25},
      xmin=1.4,
      xmax=1.46,
      xtick={1.40,1.42,1.44,1.46},
      ymin=0.42,
      ymax=0.74,
      ytick={0.5,0.6,0.7},
      legend style={
          font=\footnotesize,
          at={($(0.5,0.9)$)},
          legend columns=-1,
          anchor=center,
          /tikz/every even column/.append style={column sep=0.13cm}
      },
      legend cell align={left},
      xlabel near ticks,
      ylabel near ticks,
      cycle list name=mycolorlist_t,
      xlabel={$t$},
      ylabel={$E_{kin}$}
    ]

    % DG
    %\draw node[label={[font=\footnotesize]}] at (100,100.7) {DG};
    \addplot table [x=Time, y=INT EKIN DNS, col sep=comma, each nth point=20] {./data/ekin_perfect_les/perfectLES_DG_TauEkin.csv};
    \addlegendentry{filtered DNS}
    \addplot table [x=Time, y=INT Ekin U, col sep=comma, each nth point=20] {./data/ekin_perfect_les/perfectLES_DG_TauEkin.csv};
    \addlegendentry{perf. LES}
    \draw node[label={[font=\footnotesize]}] at (axis cs:1.45,0.642) {DG};

    % FV
    \addplot table [x=Time, y=INT EKIN DNS, col sep=comma, each nth point=20] {./data/ekin_perfect_les/perfectLES_FV_TauEkin.csv};
    \addplot table [x=Time, y=INT Ekin U, col sep=comma, each nth point=20] {./data/ekin_perfect_les/perfectLES_FV_TauEkin.csv};
    \draw node[label={[font=\footnotesize]}] at (axis cs:1.45,0.586) {FV};

    % Fourier
    \addplot table [x=Time, y=INT EKIN DNS, col sep=comma, each nth point=20] {./data/ekin_perfect_les/perfectLES_Fourier_TauEkin.csv};
    \addplot table [x=Time, y=INT Ekin U, col sep=comma, each nth point=20] {./data/ekin_perfect_les/perfectLES_Fourier_TauEkin.csv};
    \draw node[label={[font=\footnotesize]}] at (axis cs:1.45,0.464) {Fourier};

  \nextgroupplot[
      width=\figwidth,
      grid=both,
      grid style={line width=.1pt, draw=gray!5},
      major grid style={line width=.2pt,draw=gray!25},
      restrict x to domain=0:4,
      xmin=3,
      xmax=24,
      xmode=log,
      xtick={3.0,6.0,12.0,24.},
      xticklabel=\pgfmathparse{exp(\tick)}\pgfmathprintnumber{\pgfmathresult},
      ymin={5*10^(-3)},
      ymax={0.1},
      ymode=log,
      xlabel near ticks,
      ylabel near ticks,
      %log ticks with fixed point,
      %cycle list name=black white,
      cycle list name=mycolorlist,
      legend pos=south west,
      legend cell align={left},
      legend style={font=\footnotesize},
      xlabel={$k$},
    ]
  \addplot table [x index=0, y index=1, skip first n=2, col sep=space, mark=none] {./data/spectra_filter_run100/Re2000_N7_64Cells_run100_DNS_EnergySpectrum_0000001.4000000.dat};
  \addlegendentry{DNS}

  \addplot table [x index=0, y index=1, skip first n=2, col sep=space, mark=square] {./data/spectra_filter_run100/Re2000_N7_64Cells_run100_x0_y0_z0_EnergySpectrum_0000001.400000000.dat};
  \addlegendentry{DG filter}

  \addplot table [x index=0, y index=1, skip first n=2, col sep=space, mark=circle] {./data/spectra_filter_run100/Re2000_N7_64Cells_run100_FV_EnergySpectrum_0000001.400000000.dat};
  \addlegendentry{FV filter}

  \addplot table [x index=0, y index=1, skip first n=2, col sep=space, mark=*] {./data/spectra_filter_run100/Re2000_N7_64Cells_run100_Fourier_kmax12_EnergySpectrum_0000001.400000000.dat};
  \addlegendentry{Fourier filter}
%
  %\addplot[black, domain=1:64, dashed] (x,{3*x^(-5/3)});
%
  \end{groupplot}
\end{tikzpicture}

%% file: tikz/fig_filter_field_solution.tikz
\def\XIndex{0}  % Columns Index of x coordinate
\def\YIndex{1}  % Columns Index of y coordinate
\def\DataA{3}   % Columns Index of 1st Data entry (U)
\def\DataB{6}   % Columns Index of 2nd Data entry (-R_DNS)
\def\DataC{9}   % Columns Index of 3rd Data entry (R_LES-R_DNS)
\def\figwidth{0.32\textwidth} % Width/Height of figure

\def\DataAMin{-2}    % Min value for data A
\def\DataAMax{+2}    % Max value for data A
\def\DataBMin{-20}   % Min value for data B
\def\DataBMax{+20}   % Max value for data B
\def\DataCMin{-60}   % Min value for data C
\def\DataCMax{+60}   % Max value for data C

\pgfplotsset{/pgfplots/colormap={jet}{rgb255(0cm)=(0,0,128)rgb255(1cm)=(0,0,255)rgb255(3cm)=(0,255,255)rgb255(5cm)=(255,255,0)rgb255(7cm)=(255,0,0)rgb255(8cm)=(128,0,0)}}

\begin{tikzpicture}[font=\small,baseline]

  \begin{groupplot}[
        group style={
          group size=4 by 2,
          horizontal sep={0.02\textwidth},
          vertical sep={0.03\textwidth}
        },
        width=\linewidth
    ]

    % DNS
    \nextgroupplot[
      title={DNS},
      axis on top,
      enlargelimits=false,
      width={\figwidth},height={\figwidth},
      ticks=none,
      point meta min=\DataAMin,
      point meta max=\DataAMax,
      colorbar,
      %colorbar sampled,
      colorbar left,
      colorbar style={
        samples=11,
        width=0.3cm,
        title={$v_1$},
        ylabel near ticks,
        title style={at={(-0.20,0.4)},anchor=south,rotate=90},
        ytick={-2,2},
        at={(-0.15,0.)},
        anchor=south,
      },
      view={0}{90}]
      \addplot graphics[xmin=0,ymin=0,xmax=6.282,ymax=6.282] {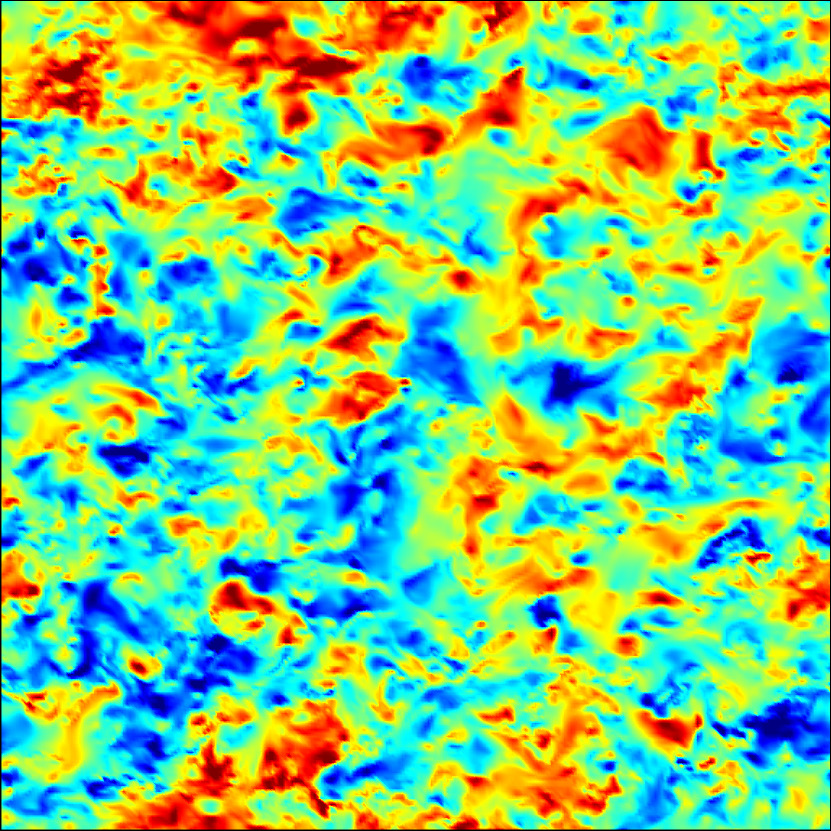};

    % DG
    \nextgroupplot[
      title={DG Filter},
      axis on top,
      enlargelimits=false,
      width={\figwidth},height={\figwidth},
      ticks=none,
      point meta min=\DataAMin,
      point meta max=\DataAMax,
      view={0}{90}]
      \addplot graphics[xmin=0,ymin=0,xmax=6.282,ymax=6.282] {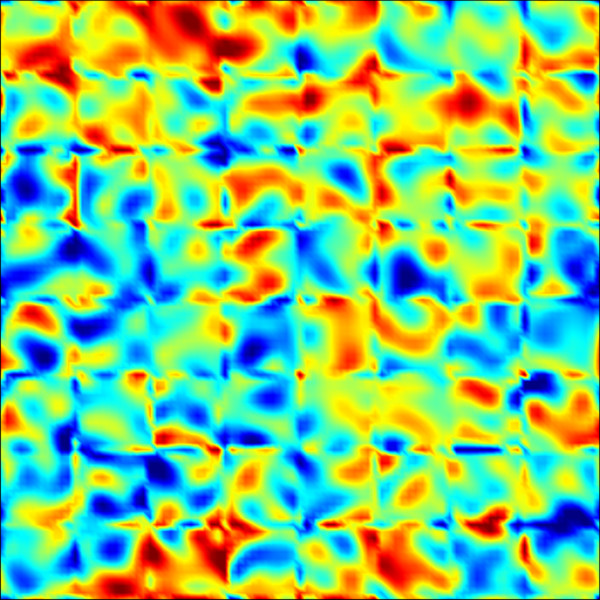};

    % FV
    \nextgroupplot[
      title={FV Filter},
      axis on top,
      enlargelimits=false,
      width={\figwidth},height={\figwidth},
      ticks=none,
      point meta min=\DataAMin,
      point meta max=\DataAMax,
      view={0}{90}]
      \addplot graphics[xmin=0,ymin=0,xmax=6.282,ymax=6.282] {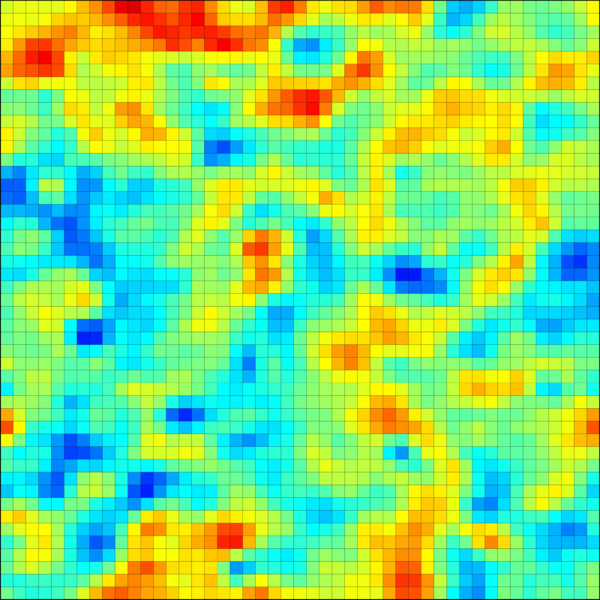};

    % Fourier
    \nextgroupplot[
      title={Fourier Filter},
      axis on top,
      enlargelimits=false,
      width={\figwidth},height={\figwidth},
      ticks=none,
      point meta min=\DataAMin,
      point meta max=\DataAMax,
      view={0}{90}]
      \addplot graphics[xmin=0,ymin=0,xmax=6.282,ymax=6.282] {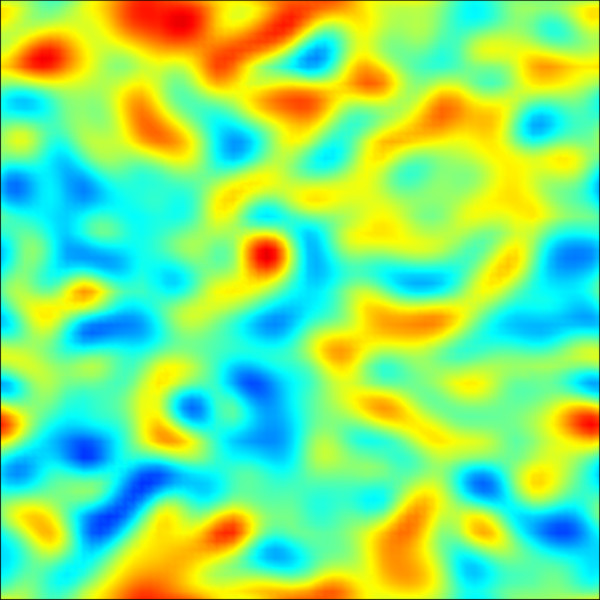};

    % ====================
    % Operator 
    % ====================

    % DNS
    \nextgroupplot[
      axis on top,
      enlargelimits=false,
      width={\figwidth},height={\figwidth},
      ticks=none,
      point meta min=\DataBMin,
      point meta max=\DataBMax,
      colorbar,
      %colorbar sampled,
      colorbar left,
      colorbar style={
        samples=11,
        width=0.3cm,
        title={$R(F(U))_1$},
        ylabel near ticks,
        title style={at={(-0.20,0.4)},anchor=south,rotate=90},
        ytick={-20,20},
        at={(-0.15,0.)},
        anchor=south,
      },
      view={0}{90}]
      \addplot graphics[xmin=0,ymin=0,xmax=6.282,ymax=6.282] {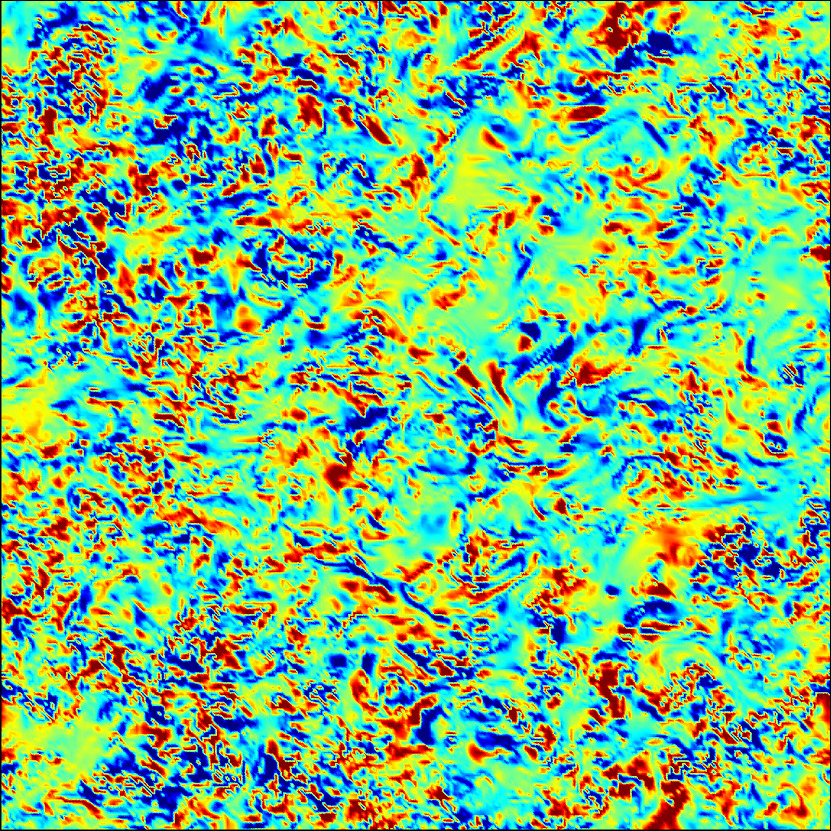};

    % DG
    \nextgroupplot[
      axis on top,
      enlargelimits=false,
      width={\figwidth},height={\figwidth},
      ticks=none,
      view={0}{90}]
      \addplot graphics[xmin=0,ymin=0,xmax=6.282,ymax=6.282] {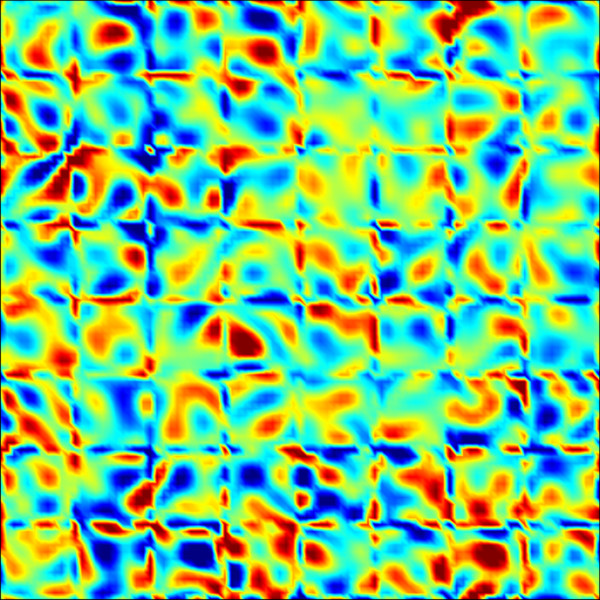};

    % FV
    \nextgroupplot[
      axis on top,
      enlargelimits=false,
      width={\figwidth},height={\figwidth},
      ticks=none,
      view={0}{90}]
      \addplot graphics[xmin=0,ymin=0,xmax=6.282,ymax=6.282] {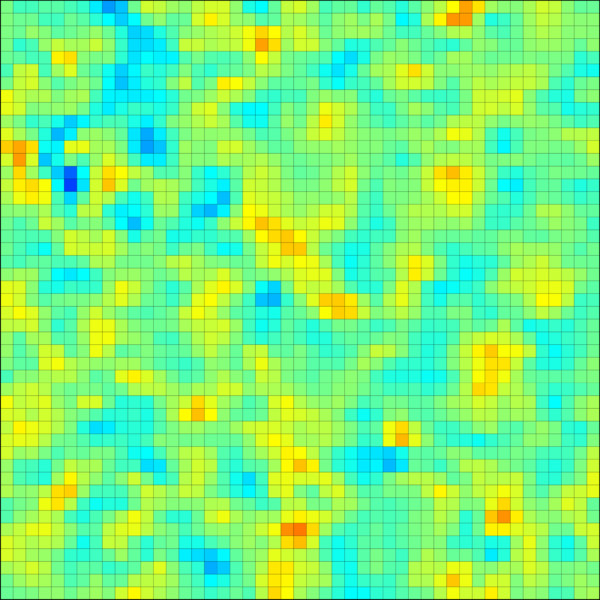};

    % Fourier
    \nextgroupplot[
      axis on top,
      enlargelimits=false,
      width={\figwidth},height={\figwidth},
      ticks=none,
      point meta min=\DataBMin,
      point meta max=\DataBMax,
      view={0}{90}]
      \addplot graphics[xmin=0,ymin=0,xmax=6.282,ymax=6.282] {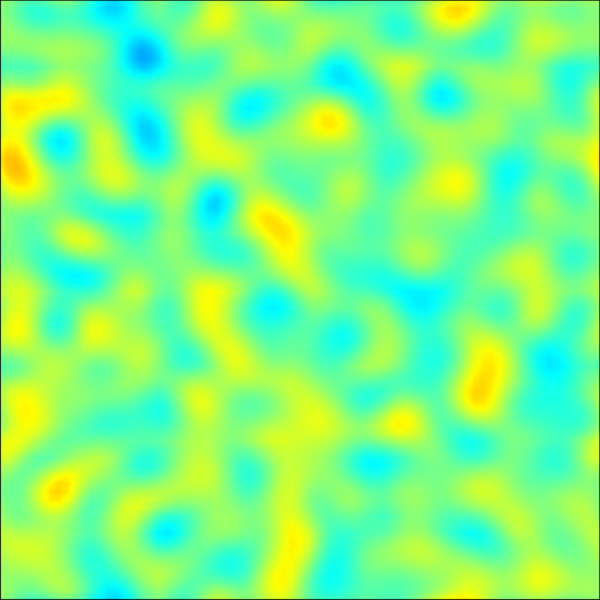};

  \end{groupplot}
\end{tikzpicture}

%% file: tikz/fig_filter_histogramm.tikz
\definecolor{histblue}{rgb}{0.0000, 0.4470, 0.7410}
\definecolor{historange}{rgb}{0.8500, 0.3250, 0.0980}

\def\histcolorA{historange}
\def\histcolorB{darkgray}
\def\histcolorC{histblue}

\begin{tikzpicture}[font=\small,baseline]
  \begin{groupplot}[
        group style={
          group size=3 by 1,
          horizontal sep={0.08\textwidth},
          vertical sep={0.05\textwidth}
        },
        xtick align=inside,
        width=0.37\linewidth,height=0.37\linewidth %height=5cm,
    ]

    % rho
    \nextgroupplot[
      grid=both,
      xmin=-20,xmax=20,
      xlabel={$(\tilde{R}(\overline{U})-\overline{R(F(U))})_0$},
      xtick={-20.,0.,20.},
      minor x tick num=1,
      minor y tick num=2,
      ytick={0.,0.9},
      ymin=0,ymax=0.9,
      ]
      \addplot[color=\histcolorA,very thick,fill, fill opacity=0.4] table[col sep=comma] {data/closure_histogramms/Re2000_N7_64Cells_run100_FV_ut1.csv};
      \addplot[color=\histcolorB,very thick,fill, fill opacity=0.2] table[col sep=comma] {data/closure_histogramms/LES_Re2000_N5_8Cells_Fourier_kmax_12_run100_ut1.csv};
      \addplot[color=\histcolorC,very thick,fill, fill opacity=0.4] table[col sep=comma] {data/closure_histogramms/Re2000_N7_64Cells_run100_x0_y0_z0_ut1.csv};

    % mom1
    \nextgroupplot[
      grid=both,
      xlabel={$(\tilde{R}(\overline{U})-\overline{R(F(U))})_1$},
      %scaled y ticks=base 10:2,
      xmin=-50,xmax=50,
      ymin=0,ymax=0.15,
      minor x tick num=1,
      minor y tick num=2,
      ytick={0.,0.15},
      xtick={-50.,0.,50.},
      legend cell align=left,
      legend style={
          /tikz/every even column/.append style={column sep=0.3cm},
          at={($(0.5,1.25)$)},
          legend columns=3,
          fill=none,
          font=\footnotesize,
          draw=black,
          anchor=center,
          align=center},
      ]
      \addplot[color=\histcolorA,very thick,fill, fill opacity=0.4] table[col sep=comma] {data/closure_histogramms/Re2000_N7_64Cells_run100_FV_ut2.csv};
      \addlegendentry{FV}
      \addplot[color=\histcolorB,very thick,fill, fill opacity=0.2] table[col sep=comma] {data/closure_histogramms/LES_Re2000_N5_8Cells_Fourier_kmax_12_run100_ut2.csv};
      \addlegendentry{Fourier}
      \addplot[color=\histcolorC,very thick,fill, fill opacity=0.4] table[col sep=comma] {data/closure_histogramms/Re2000_N7_64Cells_run100_x0_y0_z0_ut2.csv};
      \addlegendentry{DG}

    % e
    \nextgroupplot[
      grid=both,
      xlabel={$(\tilde{R}(\overline{U})-\overline{R(F(U))})_4\cdot10^{-3}$},
      xmin=-75,xmax=75,
      ymin=0,ymax=0.0002,
      minor x tick num=1,
      minor y tick num=2,
      ytick={0.,0.0002},
      xtick={-75.,0.,75.},
      xticklabels={-75,0,75},
      ]
      \addplot[color=\histcolorA,very thick,fill, fill opacity=0.4] table[col sep=comma] {data/closure_histogramms/Re2000_N7_64Cells_run100_FV_ut5.csv};
      \addplot[color=\histcolorB,very thick,fill, fill opacity=0.2] table[col sep=comma] {data/closure_histogramms/LES_Re2000_N5_8Cells_Fourier_kmax_12_run100_ut5.csv};
      \addplot[color=\histcolorC,very thick,fill, fill opacity=0.4] table[col sep=comma] {data/closure_histogramms/Re2000_N7_64Cells_run100_x0_y0_z0_ut5.csv};

  \end{groupplot}
\end{tikzpicture}

%% file: input_tex/3_ann.tex
\section{Artificial neural networks}
\label{sec:ann}
Artificial neural networks (ANN) are general function approximators, which can approximate any continuous functional relationship~$Y=f(X)$ between some input data~$X$ and output data~$Y$.
This notion is supported by the variety of universal approximation theorems for ANN in literature, e.g., \cite{Cybenko1989,Hornik1991,Lu2017}.
The wide diversity of applications for ANN gave rise to a certain richness in network architectures proposed in literature, with each network type optimized for a specific range of tasks, and a specific nature of data.
For our application, we confine ourselves to the discussion of the most basic ANN architecture, the multilayer perceptron, in \secref{sec:mlp} and of the gated recurrent unit (GRU) as a representative of recurrent neural networks (RNN) for series modelling in \secref{sec:rnn}.

\subsection{Multilayer perceptron}
\label{sec:mlp}
\begin{figure}[htb!]
  \centering
  \begin{subfigure}[c]{0.5\textwidth}
    \centering
    \includegraphics[width=\textwidth]{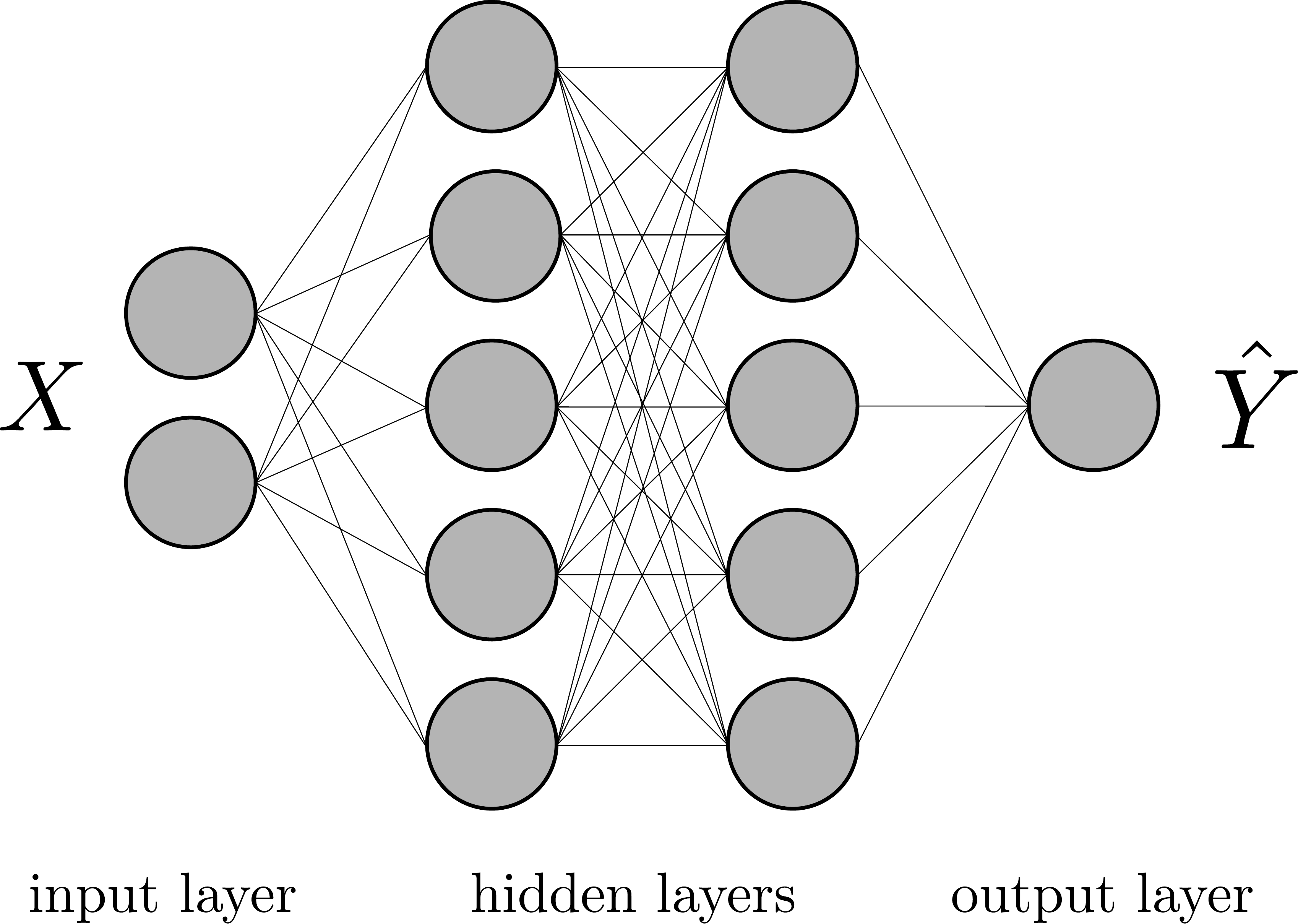}
  \end{subfigure}
  \hspace{0.05\textwidth}
  \begin{subfigure}[c]{0.4\textwidth}
    \centering
    \includegraphics[width=0.8\textwidth]{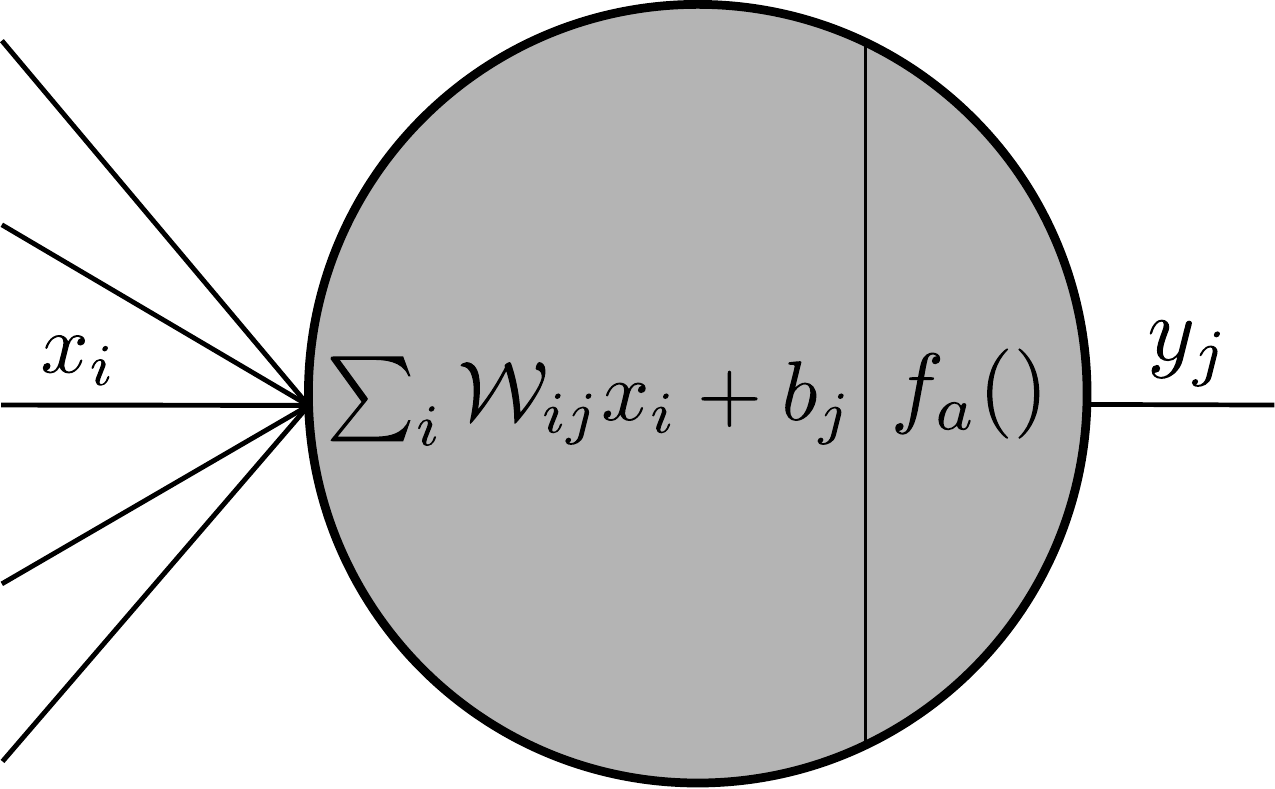}
  \end{subfigure}
  \caption{Shown on the left is the schematic architecture of an MLP with two hidden layers comprising 5 neurons respectively. Shown on the right is the layout of a single neuron with the layer index~$n$ dropped for clarity.}
  \label{fig:mlp}
\end{figure}

A multilayer perceptron (MLP) consists of multiple \textit{layers}, with each layer comprising a number of \textit{neurons}, as depicted in \figref{fig:mlp}. The layers, which are not directly connected to the input or output of the network, are called \textit{hidden layers}. Each neuron~$j$ in layer~$n$ takes the outputs of the previous layer's neurons as inputs $x^n_i$, computes the weighted sum of the inputs with the weight matrix $\mathcal{W}_{ij}^n$ and adds a bias vector $b_j^n$. Applying a non-linear \textit{activation function} $f_a$ then yields the neuron's output $y^n_j$ as
\begin{equation}
  y^n_j = f_a\left(\sum_i\mathcal{W}_{ij}^n \: x_i^n+b_j^n\right) \qquad \mathrm{with} \quad x_i^n=y^{n-1}_i \;.
\end{equation}
Common choices for the activation function are the sigmoid function ($\sigma_s=1/\left(1+e^{-x}\right)$), the hyperbolic tangent ($\sigma_t=\mathrm{tanh}\left(x\right)$), or the rectified linear unit (ReLU, $\sigma_r=\mathrm{max}\left\{x,0\right\}$).
\\
The network parameters (i.e., the weight matrices and biases) are generally initialized in a stochastic manner, e.g., with the initialization method proposed by Glorot and Bengio \cite{Glorot2010} or the method by He~et~al.~\cite{He2015}, which was proposed for networks using the ReLU activation function.
Given a dataset of training samples, where each sample consists of an input vector~$X$ and the corresponding ground truth~$Y$, the network parameters are optimized to minimize a predefined \textit{loss function}. The loss function is a measure of distance between the network's output~$\hat{Y}$ and the ground truth~$Y$ for a given training sample and thus quantifies the accuracy of the network's predictions. In the context of ANN, this optimization process is commonly referred to as \textit{training} or \textit{learning} and is achieved by computing the gradient vector of the cost function with respect to the weights, which is called \textit{backpropagation}. Training a neural network therefore refers to finding a set of network parameters, which minimizes the loss function for a given training set.
%
% GRUs
\subsection{Gated recurrent units}
\label{sec:rnn}
\begin{figure}[htb!]
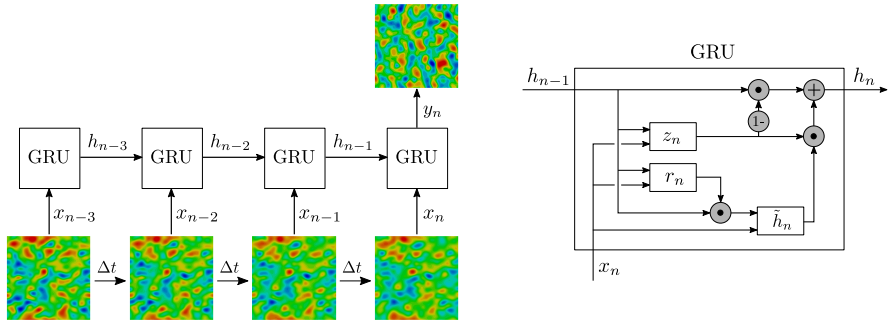

  \centering
  \begin{subfigure}[l]{0.51\textwidth}
    \includegraphics[width=\textwidth]{./fig/LSTM.eps}
  \end{subfigure}
  \hspace{0.05\textwidth}
  \begin{subfigure}[r]{0.4\textwidth}
    \includegraphics[width=\textwidth]{./fig/GRU_cell.eps}
  \end{subfigure}
  \caption{Shown on the left is the schematic of a GRU unrolled in time. At each time instant, the sample~$x_i$ and the hidden state from the previous step~$h_{i-1}$ are fed into the GRU, yielding the new state $h_i=f(h_{i-1},x_i)$. For the GRU architecture, the hidden state coincides with the layer output, i.e.,~$y_n=h_n$. The detailed architecture of a GRU cell is shown on the right. Here, all indicated operations are performed element-wise.}
  \label{fig:rnn}
\end{figure}

Recurrent neural networks (RNN) are the state-of-the-art architecture for sequential data, i.e., when the ordering of the input data is important. The underlying principle of RNN is illustrated in \figref{fig:rnn}. In addition to the current input sample~$x_n$, RNN receive their last inner state~$h_{n-1}$ as input, which allows them to retain information from previous input samples. Different types of RNN are known in literature, which differ predominantly in the way the new inner state~$h_n$ and the output~$y_n$ are computed.\\
The standard (\textit{vanilla}) RNN works like an MLP when unrolled in time. The weight matrix and bias are applied to the input (i.e., the current sample and the layer's previous output), followed by an activation function to obtain the state for the next sample. Important to note is that in contrast to an MLP, the weights of an RNN are shared for all time steps. For long input sequences, the RNN concept leads to very deep networks in time, which can cause the gradients to vanish or explode during backpropagation, as is discussed in~\cite{Bengio1994}.\\
More sophisticated RNN architectures alleviate this problem by introducing a gating mechanism, which determines the flow of information through the cell in the forward-pass and thus also the propagation of errors during backpropagation. Common representatives of this approach are the long short-term memory (LSTM) architecture proposed by Hochreiter and Schmidhuber in~\cite{Hochreiter1997} and the gated recurrent unit (GRU) proposed by Cho et al.~\cite{Cho2014}.
\\
The layout of a GRU cell is shown in \figref{fig:rnn}. The GRU receives two inputs, the current sample~$x_n$ and the previous state of the cell~$h_{n-1}$. The new state~$h_n$ is an affine combination of the old state~$h_{n-1}$ and a state candidate~$\tilde{h}_{n}$ with
\begin{equation}
  h_n = \left(1-z_n\right) \odot h_{n-1} + z_n \odot \tilde{h}_n \;.
\end{equation}
Here, $\odot$ indicates element-wise multiplication.
Important to note is that for the GRU, the inner state coincides with the layer output, i.e.,~$h_n=y_n$.
This is the major difference between the GRU and the LSTM architecture, since the latter discriminates between the hidden state and the output.
The coefficient $z_n$ is determined by the \textit{update gate}, which computes
\begin{equation}
  z_n = \sigma_s\left(\mathcal{W}_z x_n + \mathcal{U}_z h_{n-1}+b_z\right) \;,
  \label{eq:update_gate}
\end{equation}
with $\{\mathcal{W}_z,\mathcal{U}_z,b_z\}$ as trainable parameters, which are determined during the optimization process. The sigmoid function $\sigma_s$ ensures $z_n\in\left[0,1\right]$.
The state candidate $\tilde{h}_{n}$ is given by
\begin{equation}
  \tilde{h}_n = \sigma_t\left(\mathcal{W}_h x_n + \mathcal{U}_h \left(r_n\odot h_{n-1}\right)+b_z\right) \;,
\end{equation}
with $\sigma_t$ as the hyperbolic tangent and $\{\mathcal{W}_h,\mathcal{U}_h,b_h\}$ as trainable parameters. The \textit{reset gate} computes $r_n$, which determines how much of the previous state is incorporated into the new state candidate. Analogously to \eqref{eq:update_gate}, for the reset gate follows
\begin{equation}
  r_n = \sigma_s\left(\mathcal{W}_r x_n + \mathcal{U}_r h_{n-1}+b_r\right) \;.
\end{equation}
Throughout this work, we will restrict ourselves to GRU networks, since they showed to be easier to train and consistently outperformed the LSTM architecture in terms of accuracy for our application.

\section{Training the networks}
\label{sec:training}
In the following experiments, we use ANN with MLP and GRU architecture to learn the underlying non-linear mapping between the coarse-scale velocity vector $\left[\overline{v}_1,\overline{v}_2,\overline{v}_3\right]^T$ as input and the filtered DNS operator $\overline{R(F(U))}_{1,2,3}$ as the target quantity, since this is the only unknown contribution to the closure terms.
The MLP networks additionally receive the LES operator~$\tilde{R}(\overline{U})_{1,2,3}$ of the momentum equations as input, which improved the prediction accuracy considerably, as was also reported for convolutional networks in~\cite{Beck2019}.
The input vector for the MLP is therefore six-dimensional, containing the three velocity components as well as the three components of the LES operator.
For the GRU network however, including the LES operator to the input vector only resulted in minor improvements, which did not justify the substantial increase in memory consumption.
The GRU and MLP networks receive and predict only pointwise data.
For all considered networks, the input features were scaled to zero mean and unit variance on the training set.\\
To obtain training data, 10~DNS runs of the DHIT test case with different initial conditions, drawn from the same distribution, were computed.
Thereof 9~runs were used for training and validation, while a single run was kept hidden as test set to evaluate the models' performance on unseen data.
The training data was obtained by sampling the simulation results between $t=1.0$ and $t=1.9$ with an interval of $\Delta t=0.1$.
The isotropy of the flow can be exploited for data augmentation by cyclic interchange of the velocity components~$v_1,v_2,v_3$ in the different spatial directions. This allows to triple the amount of training data to about 30~million point-wise training samples for each LES filter. All networks are trained and evaluated separately for each particular LES filter.
\begin{figure}[htb!]
  \centering
  \includegraphics[width=0.8\textwidth]{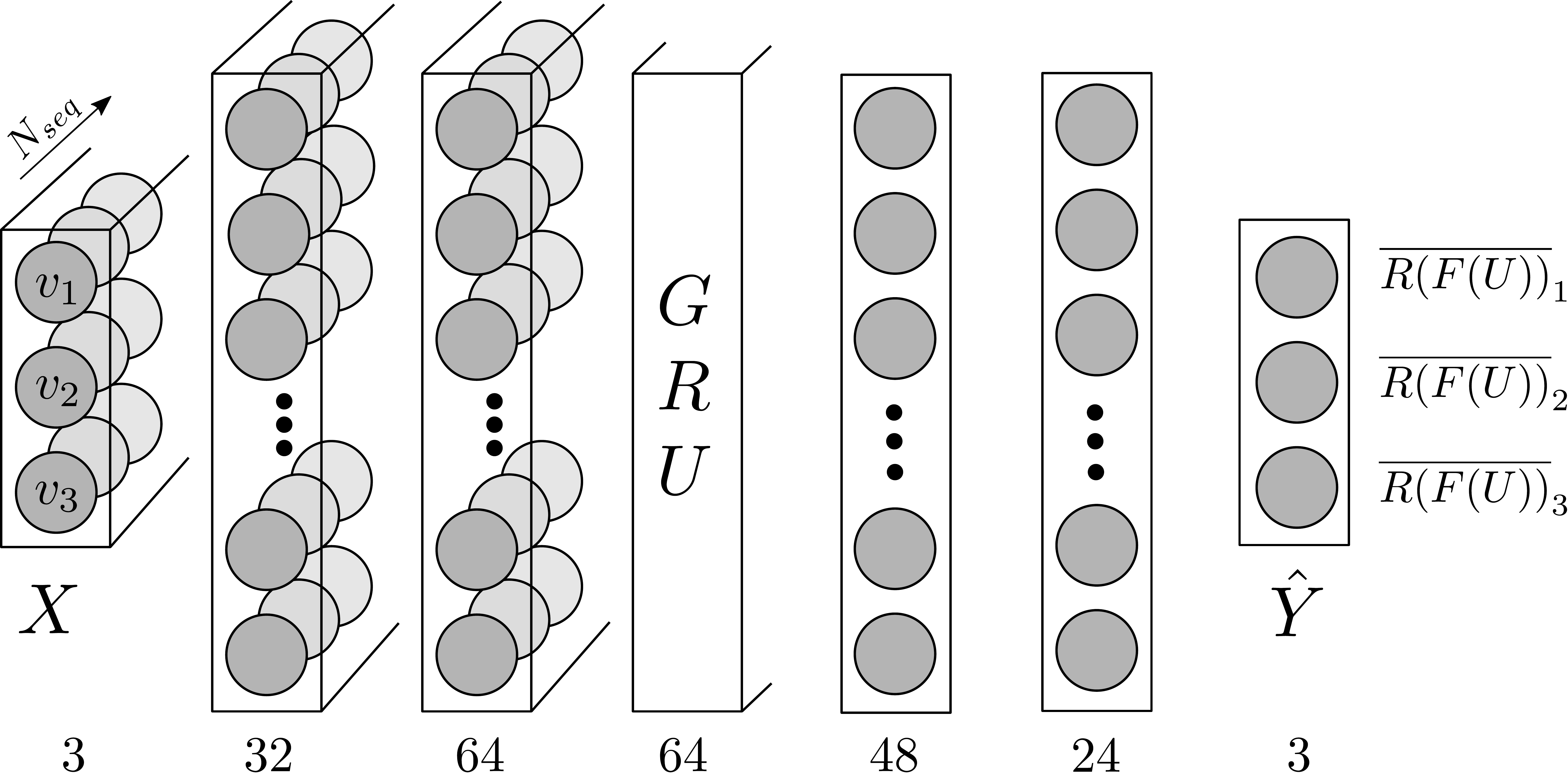}
  \caption{Architecture of the GRU network. The network consists of 5 hidden layers, with the number of neurons denoted below each layer. The first two hidden layers are executed independently for each sample in the sequence, but with the same weights. Using the \textit{many-to-one} prediction mode, the GRU layer computes a single output vector from the received inputs, as also shown in \figref{fig:rnn}.}
  \label{fig:gru_training}
\end{figure}\\
The networks investigated in this work follow the sequential architecture of \figref{fig:gru_training} with 5~hidden layers. The first two layers are dense layers with 32 and 64 neurons, followed by a GRU layer with 64 cells and again two dense layers with 48 and 24 neurons, respectively. For the MLP network, the GRU layer is replaced by a dense layer comprising 64 neurons.
All layers except the GRU layer employ ReLU activation functions and are initialized with the method proposed by He et al. \cite{He2016}.\\
For the GRU network, also the influence of different sequence lengths and sampling frequencies for the input data is investigated, with an overview of the different parameter combinations given in \tabref{tab:gru_input}. The time increment between input samples was deliberately chosen in the order of magnitude of a stable numerical timestep for the LES computation, which is around~$\Delta t\approx10^{-4}$, in order to examine the feasibility of the described approach for a priori predictions in practical LES. Thus, the predictions of the GRU networks are based on short-time dynamics on the order of the smallest resolved feature.
\begin{table}[htb!]
  \caption{Characteristics of the different input sequences for the GRU network. The total number of samples per sequence is given by~$N_{seq}$, the time increment between two samples is~$\Delta t_{seq}$, and the total time interval of the sequence, i.e., the time interval between the first and the last sample, is denoted as~$\Delta t_{tot}$.}
  \centering
  \setlength{\tabcolsep}{15pt}
  \begin{tabular}{lcccc}
    \hline
                     &            GRU1 &           GRU2  &           GRU3  \\
    \hline
    $N_{seq}$        &               3 &             10  &             21  \\
    $\Delta t_{seq}$ & $1\cdot10^{-3}$ & $1\cdot10^{-4}$ & $1\cdot10^{-4}$ \\
    $\Delta t_{tot}$ & $2\cdot10^{-3}$ & $9\cdot10^{-4}$ & $2\cdot10^{-3}$ \\
    \hline
  \end{tabular}
  \label{tab:gru_input}
\end{table}\\
The mean squared error is used as loss function for all networks and Pearson's correlation coefficient is used as additional performance metric. The correlation coefficient between two quantities $a,b$ is defined as
\begin{equation}
  \mathcal{CC}(a,b) = \frac{\mathrm{Cov(a,b)}}{\sqrt{\mathrm{Var(a)}}\sqrt{\mathrm{Var(b)}}}\;,
\end{equation}
with the covariance $\mathrm{Cov}()$ and the variance $\mathrm{Var()}$.
In addition, the mean $L_2$-error of the prediction is reported, which is obtained by integrating the prediction errors element-wise on the coarse mesh and averaging the error over all elements.\\
The whole framework was implemented in Tensorflow 2.1~\cite{Tensorflow2015} and the training was carried out on an Nvidia K40c GPU. For all networks, the Adam algorithm by Kingma and Ba~\cite{Kingma2014} was used for optimization. The networks were trained for 50 epochs with a batch size of 256 and an initial learning rate of 0.001, which was halved every 10 epochs. No additional regularization was employed during training, since the training set is sufficiently large and no overfitting was observed by comparing the losses on training and validation set.

%% file: input_tex/4_results.tex
\section{Results}
\label{sec:results}
The results are split into two parts. In \secref{sec:results_testset}, the training performance and the prediction accuracy of the networks on the test set for each LES filter are discussed. The generalization abilities of the investigated networks across different LES resolutions and filters are investigated in \secref{sec:results_generalization}. 
\subsection{Training results}
\label{sec:results_testset}
The GRU as well as the MLP architectures are able to learn the unknown closure terms from the coarse-scale input data for all three considered filter forms, whereby the GRU networks outperformed the MLP architecture.
A detailed overview of the networks' performance for the different data sets is given in \tabref{tab:results_testset} and a comparison of the predicted field solutions is shown in \figref{fig:results}.
The predicted closure terms of the GRU3 network show excellent agreement with the exact closure terms for all considered filter forms.
In addition, the GRU3 network recovers the statistical distribution of the exact closure terms almost perfectly.
The MLP architecture however cannot reproduce the exact distribution, but exhibits much lower variance in its predictions.
The MLP therefore consistently underestimates the occurring extrema of the closure terms.
This effect is most pronounced for the DG-filtered data, leading to only $24\%$ correlation between the MLP prediction and the exact terms.
\begin{figure}[htb!]
  \centering
  \input{./tikz/fig_prediction_filter.tikz}
  \caption{Shown are two-dimensional slices of the exact closure term of the x-momentum equation with the predictions of the GRU3 and the MLP architecture for the DG filter (\textit{top}), the FV filter (\textit{center}), and the Fourier filter (\textit{bottom}). The distributions of the exact and the predicted closure term are shown on the very right. The distributions are normalized to integrate to unity.}
  \label{fig:results}
\end{figure}
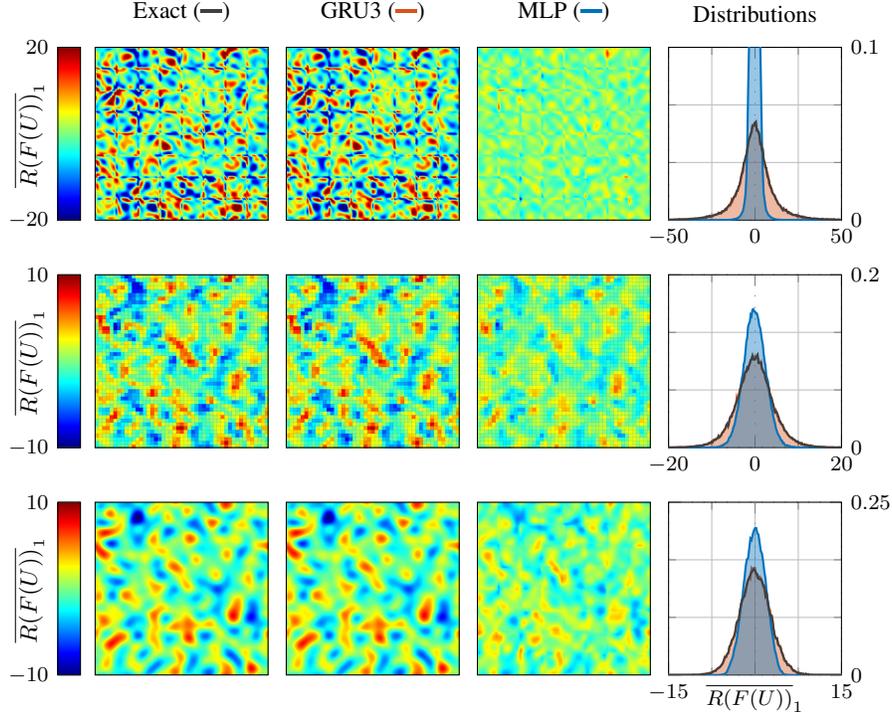\\
% GRU
The sampling of the input sequence showed to be of only minor importance for the accuracy of the GRU networks.
The GRU3 network showed the best performance for all considered datasets, since it received all 21 available timesteps and therefore the most extensive temporal information.
However, the performance of the GRU1 and GRU2 networks showed to be only slightly inferior for most considered cases.
Only for the DG-filtered data, the GRU2 showed significantly higher prediction errors than the other two networks.
A possible explanation for the similar prediction performance is that for the low Mach number test case, the numerical timestep (which was chosen as time increment in the input sequences) is small in comparison to the relevant physical time scales.
Due to the extensive temporal resolution, the larger timestep of the GRU1 is still sufficient to resolve the relevant physical phenomena and therefore to accurately predict the LES closure terms.
This finding can be exploited to reduce the temporal resolution in the input sequence during training, since storing large amounts of DNS data is prohibitive in terms of storage capacity.
\begin{table}
  \caption{Prediction accuracy of the trained ANN on the respective test sets. The GRU1 and GRU2 networks for the Fourier data were trained with halved initial learning rate, as they showed divergent behaviour otherwise. The best performance on each dataset is highlighted in bold font.}
  \centering
  \renewcommand{\arraystretch}{1.2}
  \begin{tabular}{lccccccccccc}
    \hline
          & \multicolumn{2}{c}{DG Filter} & \multicolumn{2}{c}{FV Filter} & \multicolumn{2}{c}{Fourier Filter} \\
          & $L_2$-Error &  $\mathcal{CC}$ & $L_2$-Error &  $\mathcal{CC}$ & $L_2$-Error & $\mathcal{CC}$ \\
    \hline
    MLP   &           $2.56\cdot10^{+2}$ &         $0.2456$  &         $1.02\cdot10^{+2}$ &         $0.5744$  &         $3.08\cdot10^{+1}$ &         $0.6712$  \\
    GRU1  &           $9.44\cdot10^{-1}$ & $\mathbf{0.9989}$ &         $2.53\cdot10^{-1}$ & $\mathbf{0.9992}$ &         $1.09\cdot10^{-1}$ &         $0.9992$  \\
    GRU2  &           $1.07\cdot10^{+2}$ &         $0.8157$  &         $2.33\cdot10^{-1}$ & $\mathbf{0.9992}$ &         $9.25\cdot10^{-2}$ & $\mathbf{0.9993}$ \\
    GRU3  &$\mathbf{9.14\cdot10^{-1}}$& $\mathbf{0.9989}$ & $\mathbf{2.31\cdot10^{-1}}$& $\mathbf{0.9992}$ & $\mathbf{9.15\cdot10^{-2}}$& $\mathbf{0.9993}$ \\
    \hline
  \end{tabular}
  \label{tab:results_testset}
\end{table}\\
Further experiments indicated that the prediction accuracy of the GRU networks could be improved further by increasing the network depth with addtional GRU layers.
However, the improvements with each additional layer diminished quickly while the computational effort increased. Moreover, the networks were deliberately chosen small with only around 32,000~parameters to allow their efficient evaluation for potential applications in data-driven closure models for practical LES.\\
To address the importance of closing the mass and energy equation for the DG data, which was discussed in \rmkref{rmk:closure_mass_energy}, it was investigated whether the GRU networks can also predict the closure terms for the full equation system.
To this end, the input vector for the GRU networks was extended by the density $\rho$ and the energy $e$. After training the network, it was able to predict the whole closure vector with all 5 entries up to a similar degree of accuracy, as is reported here for the momentum equation.\\
The results presented in this section are rather remarkable. They indicate that for this canonical turbulent flow, the \emph{exact} closure terms can be predicted by a recurrent neural network with (likely almost) \emph{arbitrary accuracy} for a range of different filter forms. The achieved error norms and cross correlations vastly outperform previously reported data, see e.g.~\cite{Beck2019} for results with convolutional neural networks. It should also be stressed that the exact evaluation of the closure terms requires knowledge of the full fine scale solution $U$, while the ANN-based approaches predict from the large scale quantities $\bar{U}$ only. This astonishing capability of the method used, which in a sense seems to entail some form of implicit filter inversion, will be investigated further in the future.
\subsection{Generalization}
\label{sec:results_generalization}
To test the generalization abilities of the networks, the trained GRU3 networks from \secref{sec:results_testset}, which were each trained for one particular LES filter, are applied to the test data of the other filter functions, which they have not seen during training.
The results given in \tabref{tab:results_filter} indicate, that all networks generalize well across the different filter forms. Especially for the FV and Fourier data, all networks still achieve over $99\%$ correlation between the predicted and the exact closure terms.
Only for the DG-filtered data, the prediction accuracy of the network, which was trained on the Fourier data, drops significantly.
This is expected, since the distribution of the closure terms for the DG filter exhibits much larger variance than the Fourier data.
Since the network has never learned to reproduce closure terms of this magnitude, it systematically underpredicts the extrema of the DG closure terms, which leads to a less accurate prediction.
These results indicate that, despite the statistical differences between the closure terms for different LES filters, the relationship between the temporal evolution of the coarse-scale solution and the corresponding closure terms, seems to exhibit a universal character and to transfer reasonably well across the different filter forms. 
This is not surprising, since the target quantity $\overline{R(F(U))}_i$ defines the temporal evolution of the coarse-scale solution (see \eqref{eq:filtered_nse}) and thus, the inverse mapping allows to deduce the filtered flux term $\overline{R(F(U))}_i$ from the temporal evolution of the coarse-scale inputs, which is to some extent independent of the used filter function.
\begin{table}
  \caption{Prediction accuracy of the GRU3 networks for each LES filter. The rows refer to the datasets the networks were originally trained on and the columns refer to the test data the networks are evaluated on.}
  \centering
  \renewcommand{\arraystretch}{1.2}
  \begin{tabular}{lcccccc}
    \hline
          & \multicolumn{2}{c}{DG Data}  &  \multicolumn{2}{c}{FV Data}  &  \multicolumn{2}{c}{Fourier Data} \\
          & $L_2$-Error &  $\mathcal{CC}$  &  $L_2$-Error &  $\mathcal{CC}$  &  $L_2$-Error & $\mathcal{CC}$ \\
    \hline
    GRU3-DG      & $9.14\cdot10^{-1}$ & $0.9989$  &  $3.60\cdot10^{-1}$ & $0.9988$  &  $2.57\cdot10^{-1}$& $0.9979$ \\
    GRU3-FV      & $1.46\cdot10^{ 0}$ & $0.9867$  &  $2.31\cdot10^{-1}$ & $0.9992$  &  $9.49\cdot10^{-2}$& $0.9993$ \\
    GRU3-Fourier & $9.15\cdot10^{ 0}$ & $0.8888$  &  $3.79\cdot10^{-1}$ & $0.9988$  &  $9.15\cdot10^{-2}$& $0.9993$ \\
    \hline
  \end{tabular}
  \label{tab:results_filter}
\end{table}
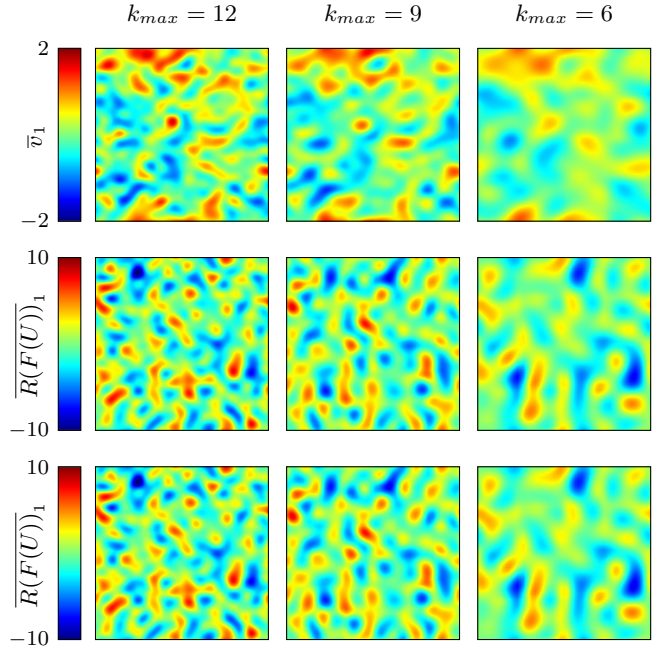
\begin{figure}[htb!]
  \centering
  \input{./tikz/fig_prediction_resolution_Fourier.tikz}
  \caption{Shown are the filtered x-velocity as network input~(\textit{top}), the exact DNS part of the closure term as network target quantity~(\textit{center}), and the predictions of a single GRU3 network for different cutoff wavenumbers $k_{max}$ of the Fourier filter~(\textit{bottom}). The network was only trained on Fourier-filtered data with~$k_{max}=12$.}
  \label{fig:results_resolution_fourier}
\end{figure}\\
In a following step, the influence of the LES filter width on the GRU3 networks' prediction accuracy is examined.
For the implicit LES filters, this corresponds to changing the resolution of the discretization.
Thus, the number of elements in each spatial direction was reduced to $75\%$ and $50\%$ of the baseline resolution, on which the networks were trained.
This yields a total of $36^3$ and $24^3$ degrees of freedom for the FV and DG filter, and to a cutoff wavenumber of $k_{max}=9$ and $k_{max}=6$ for the Fourier filter, respectively.
Due to the three-dimensional domain, the lowest resolution corresponds to only one eighth of the total degrees of freedom in comparison to the baseline resolution.
Each network is only evaluated on the LES filter it was trained on, but the original version trained on the finest grid was used, i.e. the network was not retained on the coarser grids.
The results in \tabref{tab:results_resolution} indicate that all networks are able to generalize across the different resolutions and retain low prediction errors for all examined cases.
The exact and the predicted field solution, as well as the filtered velocity field, are examplarily shown for the Fourier-filtered data in \figref{fig:results_resolution_fourier} for the different resolutions.
This ability to generalize across different resolutions can be explained by the same arguments as stated above.
\begin{table}
  \caption{Prediction performance of the GRU3 networks on LES data with different resolutions, but with the LES filter the respective networks are trained on. The resolution is quantified by the degrees of freedom (DOF) in each spatial dimension. For the Fourier filter, the baseline resolution of 48 DOF refers to $k_{max}=12$ and the lower resolutions to $k_{max}=9$ and $k_{max}=6$, respectively.}
  \centering
  \renewcommand{\arraystretch}{1.2}
  \begin{tabular}{lcccccc}
    \hline
          & \multicolumn{2}{c}{DG Filter}  &  \multicolumn{2}{c}{FV Filter}  &  \multicolumn{2}{c}{Fourier Filter} \\
          & $L_2$-Error &  $\mathcal{CC}$  &  $L_2$-Error &  $\mathcal{CC}$  &  $L_2$-Error & $\mathcal{CC}$ \\
    \hline
    48 DOF & $9.14\cdot10^{-1}$ & $0.9989$  &  $2.31\cdot10^{-1}$ & $0.9992$  &  $9.15\cdot10^{-2}$& $0.9993$ \\
    36 DOF & $1.15\cdot10^{ 0}$ & $0.9984$  &  $2.94\cdot10^{-1}$ & $0.9986$  &  $1.23\cdot10^{-1}$& $0.9989$ \\
    24 DOF & $1.10\cdot10^{ 0}$ & $0.9975$  &  $2.19\cdot10^{-1}$ & $0.9984$  &  $1.16\cdot10^{-2}$& $0.9986$ \\
    \hline
  \end{tabular}
  \label{tab:results_resolution}
\end{table}

%% file: tikz/fig_prediction_filter.tikz
\def\XIndex{0}  % Columns Index of x coordinate
\def\YIndex{1}  % Columns Index of y coordinate
\def\DataA{3}   % Columns Index of 1st Data entry (U)
\def\figwidth{0.32\textwidth} % Width/Height of figure
\def\DataDGMin{-20}    % Min value for data A
\def\DataDGMax{+20}    % Max value for data A
\def\DataFourierMin{-10}    % Min value for data A
\def\DataFourierMax{+10}    % Max value for data A
\def\DataFVMin{-10}    % Min value for data A
\def\DataFVMax{+10}    % Max value for data A

\definecolor{histblue}{rgb}{0.0000, 0.4470, 0.7410}
\definecolor{historange}{rgb}{0.8500, 0.3250, 0.0980}

\def\histcolorA{historange}
\def\histcolorC{darkgray}
\def\histcolorB{histblue}

\pgfplotsset{/pgfplots/colormap={jet}{rgb255(0cm)=(0,0,128)rgb255(1cm)=(0,0,255)rgb255(3cm)=(0,255,255)rgb255(5cm)=(255,255,0)rgb255(7cm)=(255,0,0)rgb255(8cm)=(128,0,0)}}

\begin{tikzpicture}[font=\small,baseline]

  \begin{groupplot}[
        group style={
          group size=4 by 3,
          horizontal sep={0.02\textwidth},
          vertical sep={0.06\textwidth}
        },
        width=\linewidth
    ]

    % =========================================
    % DG
    % =========================================

    % Exact - DG
    \nextgroupplot[
      title={Exact (\textcolor{\histcolorC}{\rule[2pt]{8pt}{1.2pt}})},
      axis on top,
      enlargelimits=false,
      width={\figwidth},height={\figwidth},
      ticks=none,
      point meta min=\DataDGMin,
      point meta max=\DataDGMax,
      colorbar,
      %colorbar sampled,
      colorbar left,
      colorbar style={
        samples=11,
        width=0.3cm,
        title={$\overline{R(F(U))}_1$},
        ylabel near ticks,
        title style={at={(-0.20,0.4)},anchor=south,rotate=90},
        ytick={-20,20},
        at={(-0.15,0.)},
        anchor=south,
        %height=0.9*\pgfkeysvalueof{/pgfplots/parent axis height},
      },
      view={0}{90}]
      \addplot graphics[xmin=0,ymin=0,xmax=6.282,ymax=6.282] {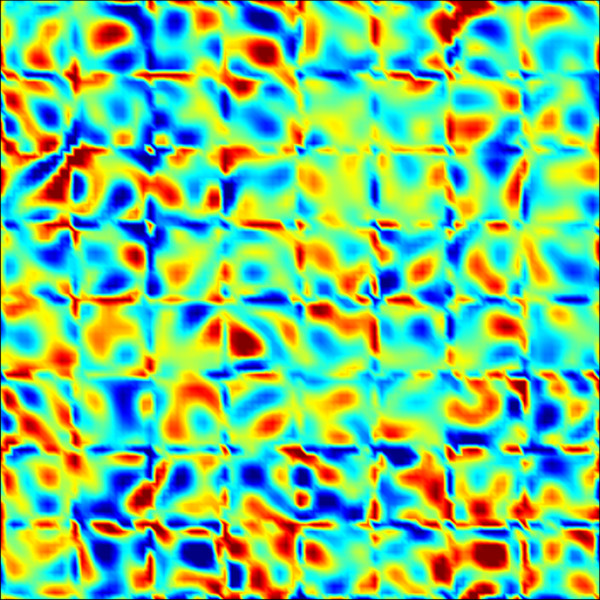};

    % GRU3 - DG
    \nextgroupplot[
      title={GRU3 (\textcolor{\histcolorA}{\rule[2pt]{8pt}{1.2pt}})},
      axis on top,
      enlargelimits=false,
      width={\figwidth},height={\figwidth},
      ticks=none,
      view={0}{90}]
      \addplot graphics[xmin=0,ymin=0,xmax=6.282,ymax=6.282] {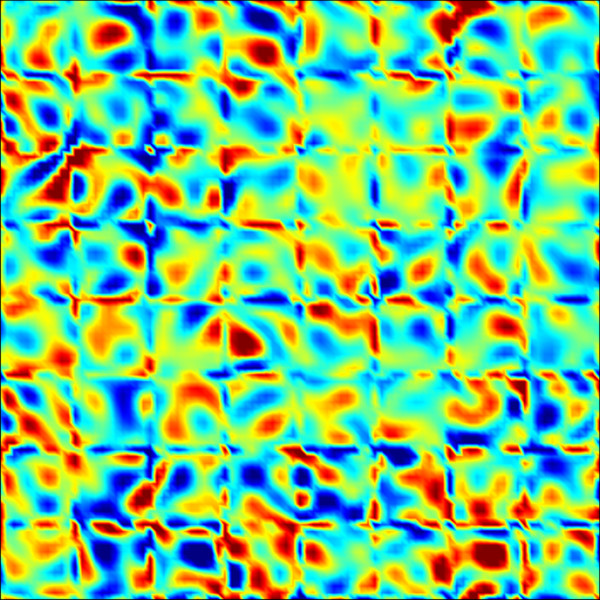};

    % MLP - DG
    \nextgroupplot[
      title={MLP (\textcolor{\histcolorB}{\rule[2pt]{8pt}{1.2pt}})},
      axis on top,
      enlargelimits=false,
      width={\figwidth},height={\figwidth},
      ticks=none,
      view={0}{90}]
      \addplot graphics[xmin=0,ymin=0,xmax=6.282,ymax=6.282] {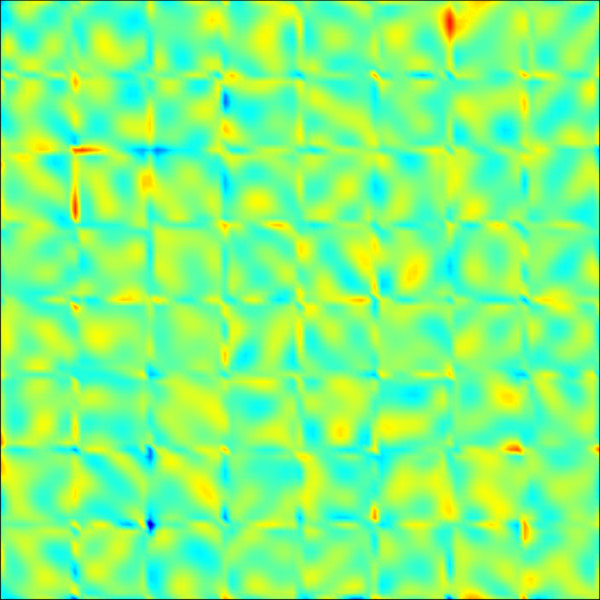};

    % Histograms
    \nextgroupplot[
      title={Distributions},
      grid=both,
      width={\figwidth},height={\figwidth},
      yticklabel pos=right,
      xmin=-50,xmax=50,
      ymin=0,ymax=0.1,
      minor x tick num=1,
      minor y tick num=2,
      ytick={0.,0.1},
      xtick={-50.,0.,50.},
      ]
      \addplot[color=\histcolorA,thick,fill, fill opacity=0.4] table[col sep=comma] {data/prediction_filter/histogram/DG_GRU3_result_dudt_2.csv};
      \addplot[color=\histcolorB,thick,fill, fill opacity=0.4] table[col sep=comma] {data/prediction_filter/histogram/DG_MLP_result_dudt_2.csv};
      \addplot[color=\histcolorC, thick] table[col sep=comma] {data/prediction_filter/histogram/Re2000_N7_64Cells_run100_x0_y0_z0_dudt_2.csv};

    % =========================================
    % FV
    % =========================================

    % Exact - FV
    \nextgroupplot[
      axis on top,
      enlargelimits=false,
      width={\figwidth},height={\figwidth},
      ticks=none,
      point meta min=\DataFVMin,
      point meta max=\DataFVMax,
      colorbar,
      %colorbar sampled,
      colorbar left,
      colorbar style={
        samples=11,
        width=0.3cm,
        title={$\overline{R(F(U))}_1$},
        ylabel near ticks,
        title style={at={(-0.20,0.4)},anchor=south,rotate=90},
        ytick={\DataFVMin,\DataFVMax},
        at={(-0.15,0.)},
        anchor=south,
        %height=0.9*\pgfkeysvalueof{/pgfplots/parent axis height},
      },
      view={0}{90}]
      \addplot graphics[xmin=0,ymin=0,xmax=6.282,ymax=6.282] {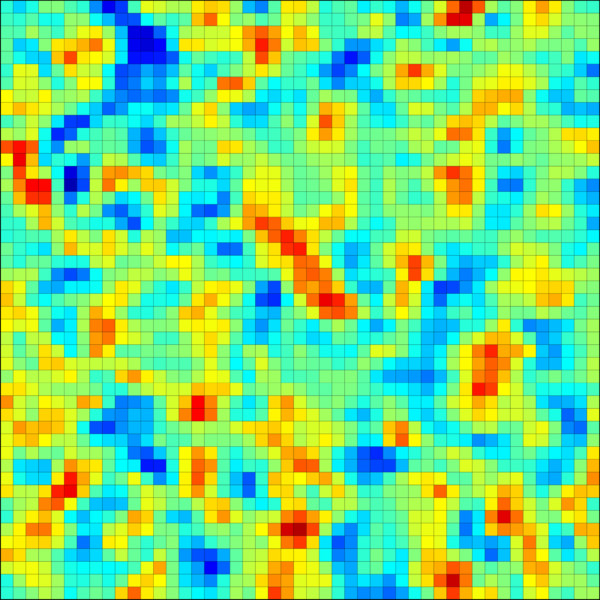};

    % GRU3 - FV
    \nextgroupplot[
      axis on top,
      enlargelimits=false,
      width={\figwidth},height={\figwidth},
      ticks=none,
      view={0}{90}]
      \addplot graphics[xmin=0,ymin=0,xmax=6.282,ymax=6.282] {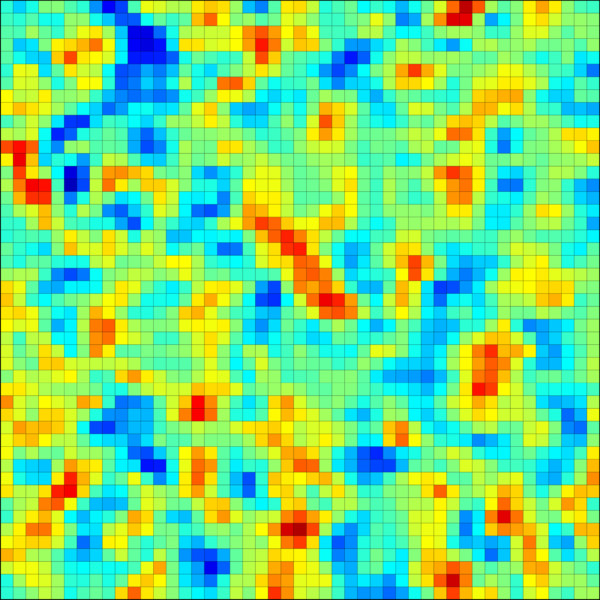};

    % MLP - FV
    \nextgroupplot[
      axis on top,
      enlargelimits=false,
      width={\figwidth},height={\figwidth},
      ticks=none,
      view={0}{90}]
      \addplot graphics[xmin=0,ymin=0,xmax=6.282,ymax=6.282] {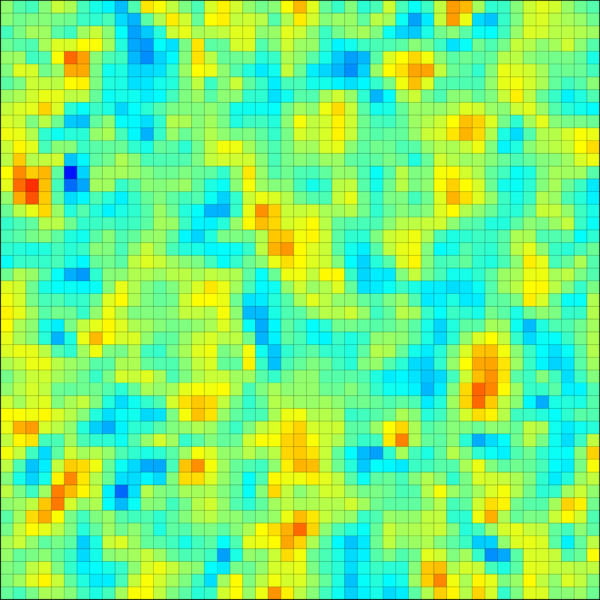};

    % Histograms
    \nextgroupplot[
      grid=both,
      width={\figwidth},height={\figwidth},
      yticklabel pos=right,
      xmin=-20,xmax=20,
      ymin=0,ymax=0.2,
      minor x tick num=1,
      minor y tick num=2,
      ytick={0.,0.2},
      xtick={-20.,0.,20.},
      ]
      \addplot[color=\histcolorA,thick,fill, fill opacity=0.4] table[col sep=comma] {data/prediction_filter/histogram/FV_GRU3_result_dudt_2.csv};
      \addplot[color=\histcolorB,thick,fill, fill opacity=0.4] table[col sep=comma] {data/prediction_filter/histogram/FV_MLP_result_dudt_2.csv};
      \addplot[color=\histcolorC,thick] table[col sep=comma] {data/prediction_filter/histogram/Re2000_N7_64Cells_run100_FV_dudt_2.csv};

    %% =========================================
    %% Fourier
    %% =========================================

    % Exact - Fourier
    \nextgroupplot[
      axis on top,
      enlargelimits=false,
      width={\figwidth},height={\figwidth},
      ticks=none,
      point meta min=\DataFourierMin,
      point meta max=\DataFourierMax,
      colorbar,
      %colorbar sampled,
      colorbar left,
      colorbar style={
        samples=11,
        width=0.3cm,
        title={$\overline{R(F(U))}_1$},
        ylabel near ticks,
        title style={at={(-0.20,0.4)},anchor=south,rotate=90},
        ytick={\DataFourierMin,\DataFourierMax},
        at={(-0.15,0.)},
        anchor=south,
        %height=0.9*\pgfkeysvalueof{/pgfplots/parent axis height},
      },
      view={0}{90}]
      \addplot graphics[xmin=0,ymin=0,xmax=6.282,ymax=6.282] {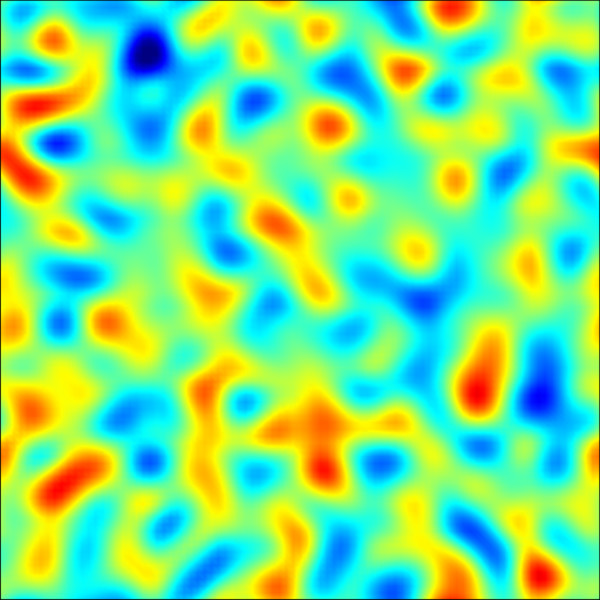};

    % GRU3 - Fourier
    \nextgroupplot[
      axis on top,
      enlargelimits=false,
      width={\figwidth},height={\figwidth},
      ticks=none,
      view={0}{90}]
      \addplot graphics[xmin=0,ymin=0,xmax=6.282,ymax=6.282] {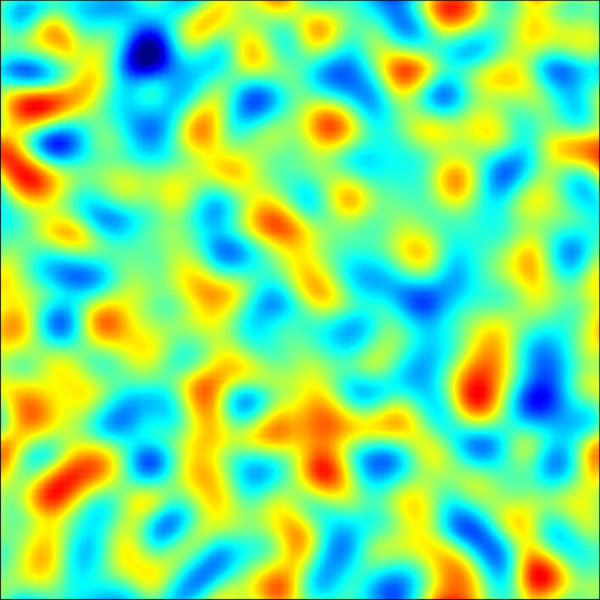};

    % MLP - Fourier
    \nextgroupplot[
      axis on top,
      enlargelimits=false,
      width={\figwidth},height={\figwidth},
      ticks=none,
      view={0}{90}]
      \addplot graphics[xmin=0,ymin=0,xmax=6.282,ymax=6.282] {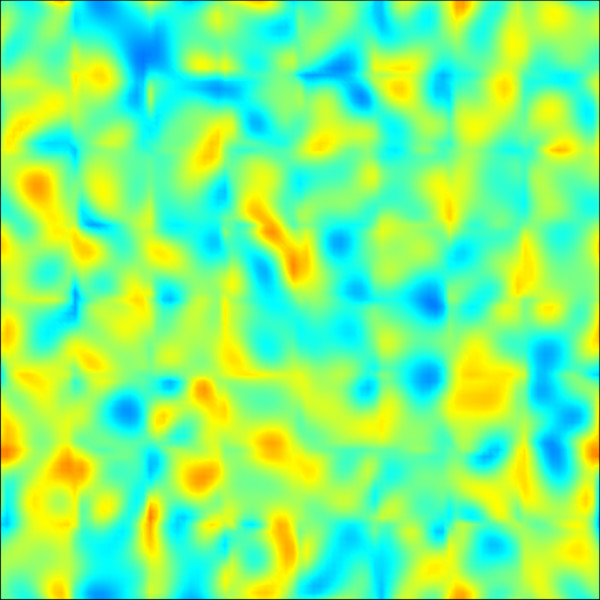};

    % Histograms
    \nextgroupplot[
      grid=both,
      width={\figwidth},height={\figwidth},
      yticklabel pos=right,
      xmin=-15,xmax=15,
      ymin=0,ymax=0.25,
      extra x ticks={0.},
      extra x tick labels={$\overline{R(F(U))}_1$},
      minor x tick num=3,
      minor y tick num=2,
      ytick={0.,0.25},
      xtick={-15.,15.},
      ]
      \addplot[color=\histcolorA,thick,fill, fill opacity=0.4] table[col sep=comma] {data/prediction_filter/histogram/Fourier_GRU3_result_dudt_2.csv};
      \addplot[color=\histcolorB,thick,fill, fill opacity=0.4] table[col sep=comma] {data/prediction_filter/histogram/Fourier_MLP_result_dudt_2.csv};
      \addplot[color=\histcolorC, thick] table[col sep=comma] {data/prediction_filter/histogram/Re2000_N7_64Cells_run100_x0_y0_z0_filtered_dudt_2.csv};

  \end{groupplot}
\end{tikzpicture}

%% file: tikz/fig_prediction_resolution_Fourier.tikz
\def\XIndex{0}  % Columns Index of x coordinate
\def\YIndex{1}  % Columns Index of y coordinate
\def\DataA{3}   % Columns Index of 1st Data entry (U)
\def\figwidth{0.32\textwidth} % Width/Height of figure
\def\DataRMin{-10}    % Min value for data A
\def\DataRMax{+10}    % Max value for data A
\def\DatavMin{-2}    % Min value for data A
\def\DatavMax{+2}    % Max value for data A

\pgfplotsset{/pgfplots/colormap={jet}{rgb255(0cm)=(0,0,128)rgb255(1cm)=(0,0,255)rgb255(3cm)=(0,255,255)rgb255(5cm)=(255,255,0)rgb255(7cm)=(255,0,0)rgb255(8cm)=(128,0,0)}}

\begin{tikzpicture}[font=\small,baseline]

  \begin{groupplot}[
        group style={
          %group size=4 by 3,
          group size=3 by 3,
          horizontal sep={0.02\textwidth},
          vertical sep={0.04\textwidth}
        },
        width=\linewidth
    ]

    % =========================================
    % V_1
    % =========================================

    % V_1 - 8 Elems
    \nextgroupplot[
      title={$k_{max}=12$},
      axis on top,
      enlargelimits=false,
      width={\figwidth},height={\figwidth},
      ticks=none,
      point meta min=\DatavMin,
      point meta max=\DatavMax,
      colorbar,
      %colorbar sampled,
      colorbar left,
      colorbar style={
        samples=11,
        width=0.3cm,
        title={$\overline{v}_1$},
        ylabel near ticks,
        title style={at={(-0.20,0.4)},anchor=south,rotate=90},
        ytick={-2,2},
        at={(-0.15,0.)},
        anchor=south,
        %height=0.9*\pgfkeysvalueof{/pgfplots/parent axis height},
      },
      view={0}{90}]
      \addplot graphics[xmin=0,ymin=0,xmax=6.282,ymax=6.282] {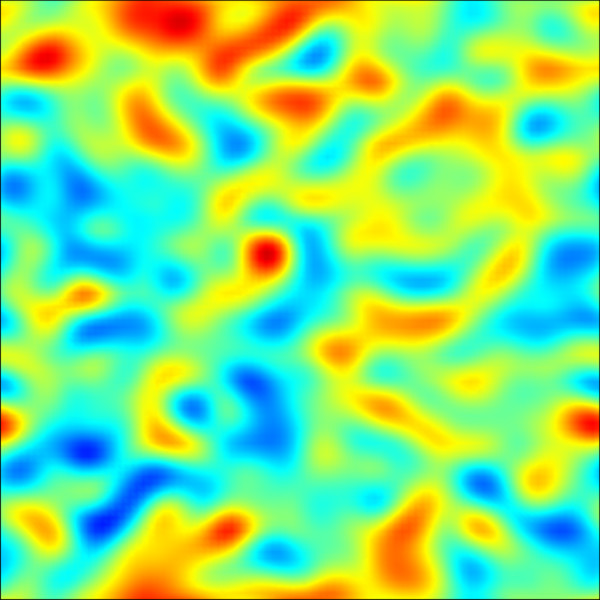};

    % V_1 - 6 Elems
    \nextgroupplot[
      title={$k_{max}=9$},
      axis on top,
      enlargelimits=false,
      width={\figwidth},height={\figwidth},
      ticks=none,
      view={0}{90}]
      \addplot graphics[xmin=0,ymin=0,xmax=6.282,ymax=6.282] {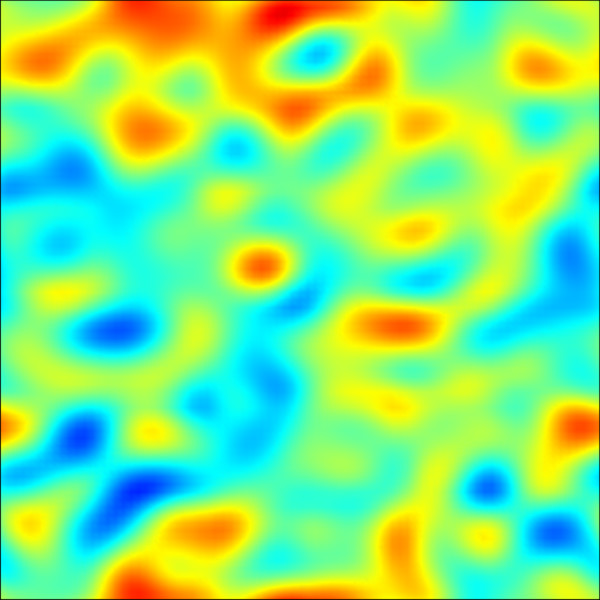};

    % V_1 - 4 Elems
    \nextgroupplot[
      title={$k_{max}=6$},
      axis on top,
      enlargelimits=false,
      width={\figwidth},height={\figwidth},
      ticks=none,
      view={0}{90}]
      \addplot graphics[xmin=0,ymin=0,xmax=6.282,ymax=6.282] {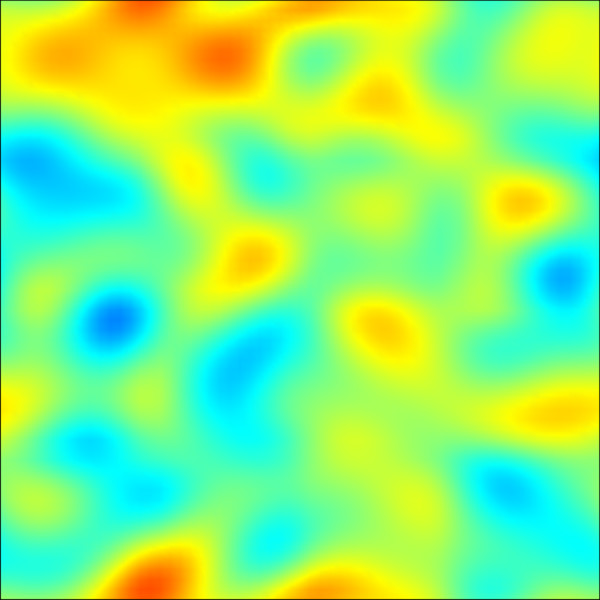};

    % =========================================
    % R_1 exact
    % =========================================

    % R_1 - 8 Elems
    \nextgroupplot[
      axis on top,
      enlargelimits=false,
      width={\figwidth},height={\figwidth},
      ticks=none,
      point meta min=\DataRMin,
      point meta max=\DataRMax,
      colorbar,
      %colorbar sampled,
      colorbar left,
      colorbar style={
        samples=11,
        width=0.3cm,
        title={$\overline{R(F(U))}_1$},
        ylabel near ticks,
        title style={at={(-0.20,0.4)},anchor=south,rotate=90},
        ytick={-10,10},
        at={(-0.15,0.)},
        anchor=south,
        %height=0.9*\pgfkeysvalueof{/pgfplots/parent axis height},
      },
      view={0}{90}]
      \addplot graphics[xmin=0,ymin=0,xmax=6.282,ymax=6.282] {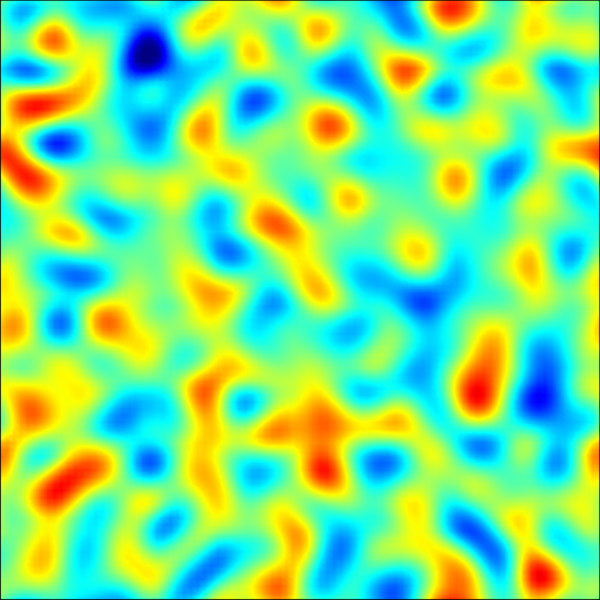};

    % R_1 exact - 6 Elems
    \nextgroupplot[
      axis on top,
      enlargelimits=false,
      width={\figwidth},height={\figwidth},
      ticks=none,
      view={0}{90}]
      \addplot graphics[xmin=0,ymin=0,xmax=6.282,ymax=6.282] {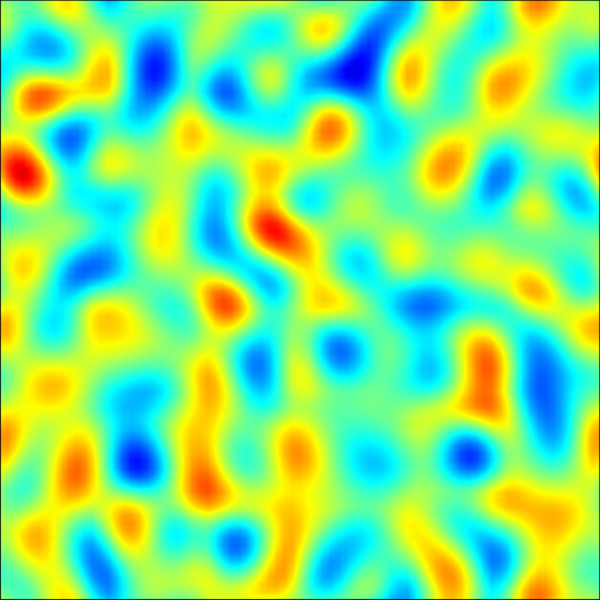};

    % R_1 exact - 4 Elems
    \nextgroupplot[
      axis on top,
      enlargelimits=false,
      width={\figwidth},height={\figwidth},
      ticks=none,
      view={0}{90}]
      \addplot graphics[xmin=0,ymin=0,xmax=6.282,ymax=6.282] {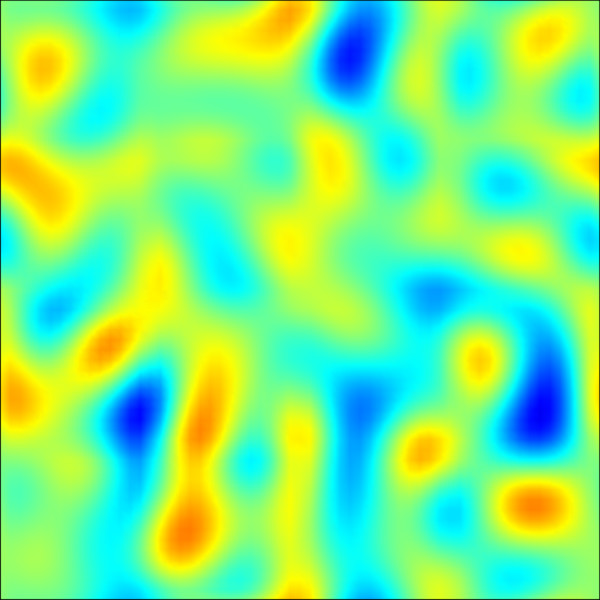};

    % =========================================
    % R_1 GRU3
    % =========================================

    % R_1 - 8 Elems
    \nextgroupplot[
      axis on top,
      enlargelimits=false,
      width={\figwidth},height={\figwidth},
      ticks=none,
      point meta min=\DataRMin,
      point meta max=\DataRMax,
      colorbar,
      %colorbar sampled,
      colorbar left,
      colorbar style={
        samples=11,
        width=0.3cm,
        title={$\overline{R(F(U))}_1$},
        ylabel near ticks,
        title style={at={(-0.20,0.4)},anchor=south,rotate=90},
        ytick={-10,10},
        at={(-0.15,0.)},
        anchor=south,
        %height=0.9*\pgfkeysvalueof{/pgfplots/parent axis height},
      },
      view={0}{90}]
      \addplot graphics[xmin=0,ymin=0,xmax=6.282,ymax=6.282] {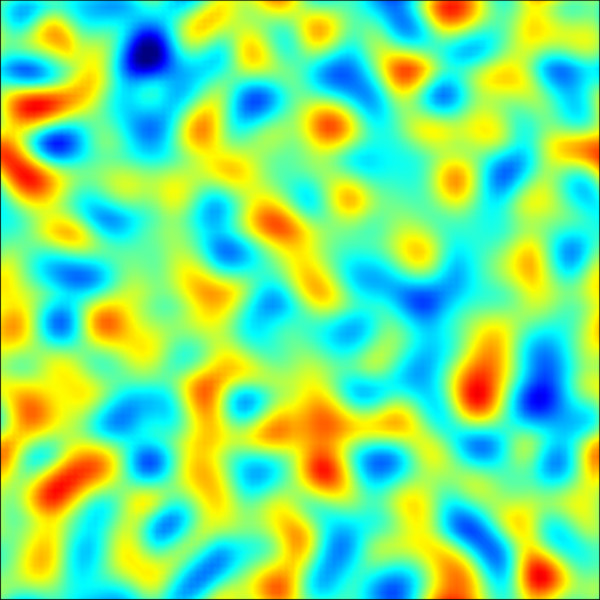};

    % R_1 GRU3 - 6 Elems
    \nextgroupplot[
      axis on top,
      enlargelimits=false,
      width={\figwidth},height={\figwidth},
      ticks=none,
      view={0}{90}]
      \addplot graphics[xmin=0,ymin=0,xmax=6.282,ymax=6.282] {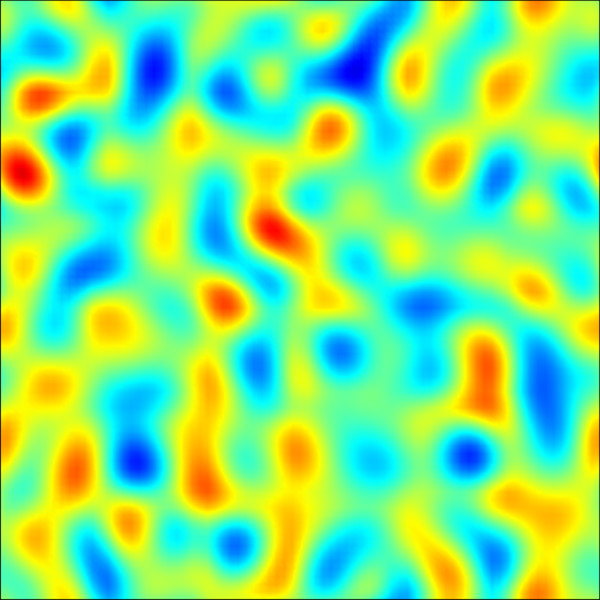};

    % R_1 GRU3 - 4 Elems
    \nextgroupplot[
      axis on top,
      enlargelimits=false,
      width={\figwidth},height={\figwidth},
      ticks=none,
      view={0}{90}]
      \addplot graphics[xmin=0,ymin=0,xmax=6.282,ymax=6.282] {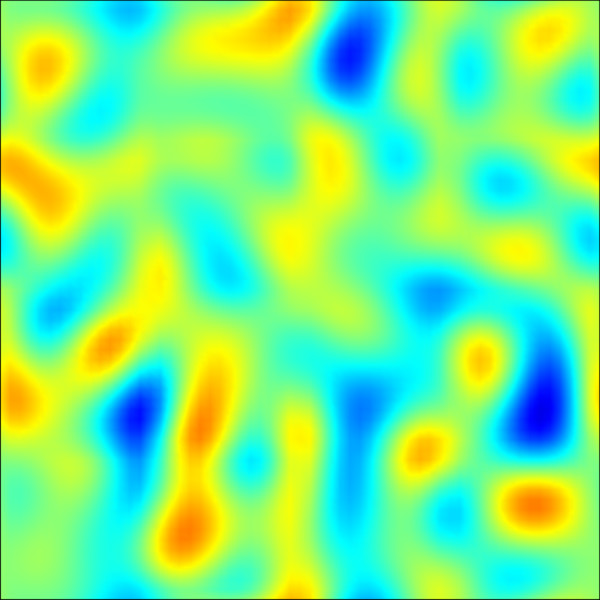};
  \end{groupplot}
\end{tikzpicture}

%% file: input_tex/5_conclusion.tex
\section{Conclusions}
\label{sec:conclusion}
In this work, we have derived a general framework for LES closure models, with the emphasis put on the effects and influences of the discretization operator. In practical LES, the discretization operator itself acts as the scale-separating filter. Its properties (non-linearity, inhomogeneity) can make a rigorous analysis tough, and often lead to the fact that the associated errors (often labelled commutation errors) are ignored or overlooked. As a result of this, closure model development is in some cases "blind" towards the underlying discretization and its induced filter form. This can be problematic, as the optimal closure terms that should govern model development are indeed a function of the filter. \\
We have investigated the dependence of the closure on the filter functions through the case of decaying homogeneous isotropic turbulence (DHIT) which serves as the simplest canonical test case for turbulence. We have build a DNS database from an ensemble of runs and have computed the exact closure terms for several LES filter functions.
For this, we used filter functions which mimic the action of a discretization filter. They are derived from the solution representation of DG and FV schemes. A third, global Fourier cutoff filter represents the explicitly filtered LES approach.
We have demonstrated that the computed closure terms are consistent in the sense that they are able to recover the filtered DNS solution in each time step, thus forming the perfect LES model.
Different neural networks with MLP and GRU architecture have been trained to predict the unknown contributions of the closure terms for each examined filter function solely from known coarse-scale data.
We have demonstrated that the GRU networks are able to approximate the exact closure terms up to a very high degree of accuracy for all considered filter forms.
The GRU networks have also been shown to generalize well across different filter functions and filter widths, which they have not seen during training.\\
These highly accurate predictions and the networks' generalization abilities are encouraging results and mark a starting point for the development of universal data-driven LES closure models. Being able to predict the exact closure term that perfectly "fits" both the physical closure problem as well as the discretization effects offers the chance to guide discretization-adapted model development. 
Future work will be focused on applying the network predictions as closure model during practical LES.
Since direct incorporation of the network predictions might cause instabilities, as was found in \cite{Beck2019}, also a strategy to incorporate the ANN predictions into a stable data-driven model have to be examined.
Further, the application and extension of the described metholodogy to other canonical turbulent flow problems, and especially wall-bounded flows, will be investigated.